\documentclass[
  superscriptaddress,
  reprint, twocolumn,
  amsmath,amssymb,
  aps, physrev,
]{revtex4-2}

\usepackage{graphicx}
\usepackage{dcolumn}
\usepackage{bm}
\usepackage{placeins}
\usepackage{float}
\usepackage{booktabs}
\usepackage[table,dvipsnames]{xcolor}
\usepackage{soul}
\usepackage[]{markdown}
\usepackage{svg}
\usepackage{nameauth}
\usepackage{siunitx}
\DeclareSIUnit\angstromunit{\text{\AA}}
\DeclareSIUnit\angstrom{\protect \text{Å}}
\usepackage{booktabs}
\usepackage{tabularx}
\usepackage{array}
\usepackage{pdfpages}
\usepackage{pgf,pgffor}
\usepackage{hyperref}
\hypersetup{
  colorlinks=true,
  citecolor=blue,
  linkcolor=blue,
  filecolor=blue,
  urlcolor=blue,
  pdfstartview=FitH,
  pdfpagemode=UseNone
}
\usepackage{cleveref}

\newcolumntype{C}{>{\centering\arraybackslash}X}
\definecolor{errorlow}{RGB}{183,222,183}
\definecolor{errormid}{RGB}{250,221,145}
\definecolor{errorhigh}{RGB}{238,145,145}
\newcommand{\heatthreshold}[2]{%
    \pgfmathsetmacro{\score}{100*min(1,max(0,(#1)/(#2)))}%
    \pgfmathtruncatemacro{\lowerhalf}{\score < 50}%
    \ifnum\lowerhalf=1
        \pgfmathsetmacro{\mixvalue}{2*\score}%
        \edef\cellcolorname{errormid!\mixvalue!errorlow}%
    \else
        \pgfmathsetmacro{\mixvalue}{2*(\score-50)}%
        \edef\cellcolorname{errorhigh!\mixvalue!errormid}%
    \fi
    \expandafter\cellcolor\expandafter{\cellcolorname}#1%
}
\newcommand{\heatJ}[1]{\heatthreshold{#1}{1.50}}
\newcommand{\heatP}[1]{\heatthreshold{#1}{25}}

\newcommand{\orcid}[1]{\href{https://orcid.org/#1}{\includegraphics[width=8pt]{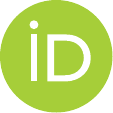}}}
\makeatletter
\AtBeginDocument{\let\LS@rot\@undefined}
\makeatother

\begin{document}
\title{
Cross-Geometry Transferability Assessment of Universal Machine Learning \\Interatomic Potentials: From Bulk Materials to Atomic Nanowires
}

\author{Pedro H. M. Zanineli \orcid{0009-0008-2359-5218}} 
\affiliation{Brazilian Nanotechnology National Laboratory (LNNano/CNPEM), 13083-100, Campinas, SP, Brazil}
\affiliation{Universidade Federal do ABC (UFABC), 09210-580, Santo André, São Paulo, Brazil}
\author{Bruno Focassio \orcid{0000-0003-4811-7729}} 
\affiliation{Brazilian Nanotechnology National Laboratory (LNNano/CNPEM), 13083-100, Campinas, SP, Brazil}
\author{Gabriel R. Schleder \orcid{0000-0003-3129-8682}}
\email[Corresponding author. ]{gabriel.schleder@lnnano.cnpem.br}
\affiliation{Brazilian Nanotechnology National Laboratory (LNNano/CNPEM), 13083-100, Campinas, SP, Brazil}
\affiliation{Universidade Federal do ABC (UFABC), 09210-580, Santo André, São Paulo, Brazil}

\begin{abstract}
Foundation machine-learning interatomic potentials (MLIPs) enable atomistic simulations at substantially lower computational cost than first-principles methods, but their reliability across structural geometries remains insufficiently understood. 
Here, we construct a density-functional-theory dataset of ZrO$_2$ configurations spanning bulk, slab, particle, neck, and atomically thin wire environments motivated by an experimentally observed ZrO$_2$ desintering process involving neck thinning and atomic wire formation. 
We first benchmark 26 pretrained MLIPs and observe pronounced geometry-dependent degradation in zero-shot predictions. 
Without any training, after only reference-energy alignment, the best zero-shot model (ORB-V3) reaches energy and force root-mean-square errors of 6~meV atom$^{-1}$ and 197.3~meV~\AA$^{-1}$, respectively, with the largest force errors in neck and wire configurations. 
We then compare zero-shot inference, fine-tuning, and training from scratch strategies. Fine-tuning yields lower energy and force errors than training from scratch, while both require comparable wall-clock time. Geometry-specific fine-tuning improves in-domain accuracy but frequently produces negative transfer to other structural classes, whereas mixed-geometry fine-tuning reduces cross-geometry errors.
Evaluations of elastic and vibrational properties, surface energies, and neck dynamics further show that rankings based on average energy and force errors do not universally predict property-level behavior. These results demonstrate that geometry-diverse target data and independent physical validations are necessary when adapting foundation MLIPs to low-coordination (ionic) nanostructures.
\end{abstract}


\maketitle

\section{Introduction}

Machine-learning interatomic potentials (MLIPs) have emerged as a practical route for extending atomistic simulations beyond the length and time scales accessible to conventional first-principles calculations \cite{schleder_dft_2019,bowler_2012,deringer_2019,zuo_2020,mishin_2021,Schleder2019}. By learning energies and forces from density-functional-theory (DFT) reference data, modern equivariant graph neural networks can approximate the underlying potential-energy landscape while reducing the computational cost of evaluating large atomic systems \cite{kozinsky_2022,batatia_2022,yang_2025}.

The increasing availability of broadly pretrained models has further motivated the development of universal (or increasingly foundation) interatomic potentials intended to transfer across compositions and structural environments. Early universal graph potentials such as M3GNet established an important baseline for broad elemental transferability by learning from large Materials Project relaxation datasets with explicit many-body graph representations \cite{m3gnet_2022}. At the same time, strictly local equivariant architectures such as Allegro provide a complementary perspective, showing that high-accuracy atomistic models can be constructed without atom-centered message passing \cite{allegro_2023}. These architectural differences are relevant for cross-geometry transfer because low-coordination motifs alter not only local chemistry but also graph connectivity and boundary structure.

Recent systematic studies of MLIP fine-tuning have shown that pretrained initialization can provide strong target-task accuracy relative to training from scratch, but also that the optimal adaptation strategy depends on the intended deployment scope: unconstrained full-parameter fine-tuning is often effective for narrow target applications, whereas replay-based strategies are better suited when preserving broader out-of-distribution robustness is required \cite{tompa_2026}.

A central unresolved question is whether such models transfer reliably across geometries \cite{Butler_6open}. Widely used pretraining datasets, including the Materials Project and Alexandria, are dominated by periodic crystalline structures \cite{jain_2013,mp_2025,schmidt_alexandria_2024}. Surfaces, nanoparticles, highly strained structures, and low-dimensional motifs are comparatively underrepresented. Consequently, a model that performs well for bulk configurations may exhibit substantially larger errors when applied to under-coordinated or non-equilibrium environments \cite{out_of_dist_2021,generalization_2024,benedini_2025,OOD_materials_2025,Deringer2026a,Deringer2026p}. This limitation is particularly important for structural processes in which several coordination regimes coexist and evolve continuously.

Zirconia provides a demanding test of this problem. Beyond its ionic character, ZrO$_2$ is also a strongly anharmonic oxide whose phase stability and finite-temperature behavior have motivated specialized machine-learned force-field studies, including on-the-fly active-learning and beyond-DFT $\Delta$-learning approaches for zirconia phase transitions \cite{verdi_zro2_2021,liu_zro2_rpa_2022,zhang_zro2_trip_2025}. High-resolution transmission electron microscopy has revealed an unusual desintering process, following previous usage for the tensile separation of initially connected monoclinic ZrO$_2$ grains through neck thinning and eventual formation of atomically thin wire-like motifs under an external stimulus \cite{focassio_rev_letter_2022,fiuza_cell_2024}. The transformation spans bulk-like regions, surfaces, particles, necks, and wire-like structures, thereby combining large changes in coordination, strain, and dimensionality.

Its ionic character also makes ZrO$_2$ a challenging system for short-range models, especially when dimensionality reduction creates surfaces, necks, and wire-like motifs with altered electrostatic screening. Recent MLIP developments address this limitation through charge-equilibration neural networks, equivariant long-range message passing, polarizable or charge-aware extensions, and latent Ewald-type electrostatic models \cite{ko_4ghdnnp_2021,lorem_2026,maruf_longrange_2025,les_2026,cheng_2025,so3lr,mace_polar}. Because the present study does not isolate the contributions of long-range electrostatics or other methodological factors, we interpret long-range effects as one possible source of residual error, potentially acting alongside force-filtering choices, trajectory-level data correlations, and incomplete optimization, rather than as a mechanism established by the present data alone.

Existing MLIP benchmarks provide essential comparisons across models, but average energy and force errors on predominantly bulk datasets do not directly establish reliability for geometry-changing processes \cite{riebesell_2025,szilvasi_2025,practical_guida_mlips}. Recent assessments further show that model reliability can be strongly property- and task-dependent, including for surfaces, phonons, finite-temperature molecular dynamics, and cross-functional transferability \cite{focassio_2024,hu_2025,liu_2025,ceder_2025}. Moreover, low test-set errors do not necessarily guarantee accurate derived properties or stable molecular dynamics \cite{Alampara2025}. A useful assessment must therefore resolve errors by geometry, distinguish initialization and adaptation strategies, and test whether improvements in supervised metrics translate to physically relevant observables.

Here, we investigate three questions. First, how strongly does zero-shot MLIP accuracy vary across bulk, slab, particle, neck, and wire ZrO$_2$ configurations? Second, under controlled optimization and data, how do zero-shot inference, fine-tuning, and training from scratch compare? Third, does exposure to multiple geometry classes improve cross-geometry generalization relative to geometry-specific adaptation? We address these questions using a geometry-diverse DFT dataset, cross-geometry learning experiments with MACE and ORB, and property-level evaluations of elastic and vibrational response, surface energies, and neck dynamics. Our working hypothesis is that pretrained initialization improves data efficiency under fixed training, but reliable transfer across the full structural process requires explicit diversity in the target-domain data.

\begin{figure*}[ht!]
    \centering
    \includegraphics[width=1\linewidth]{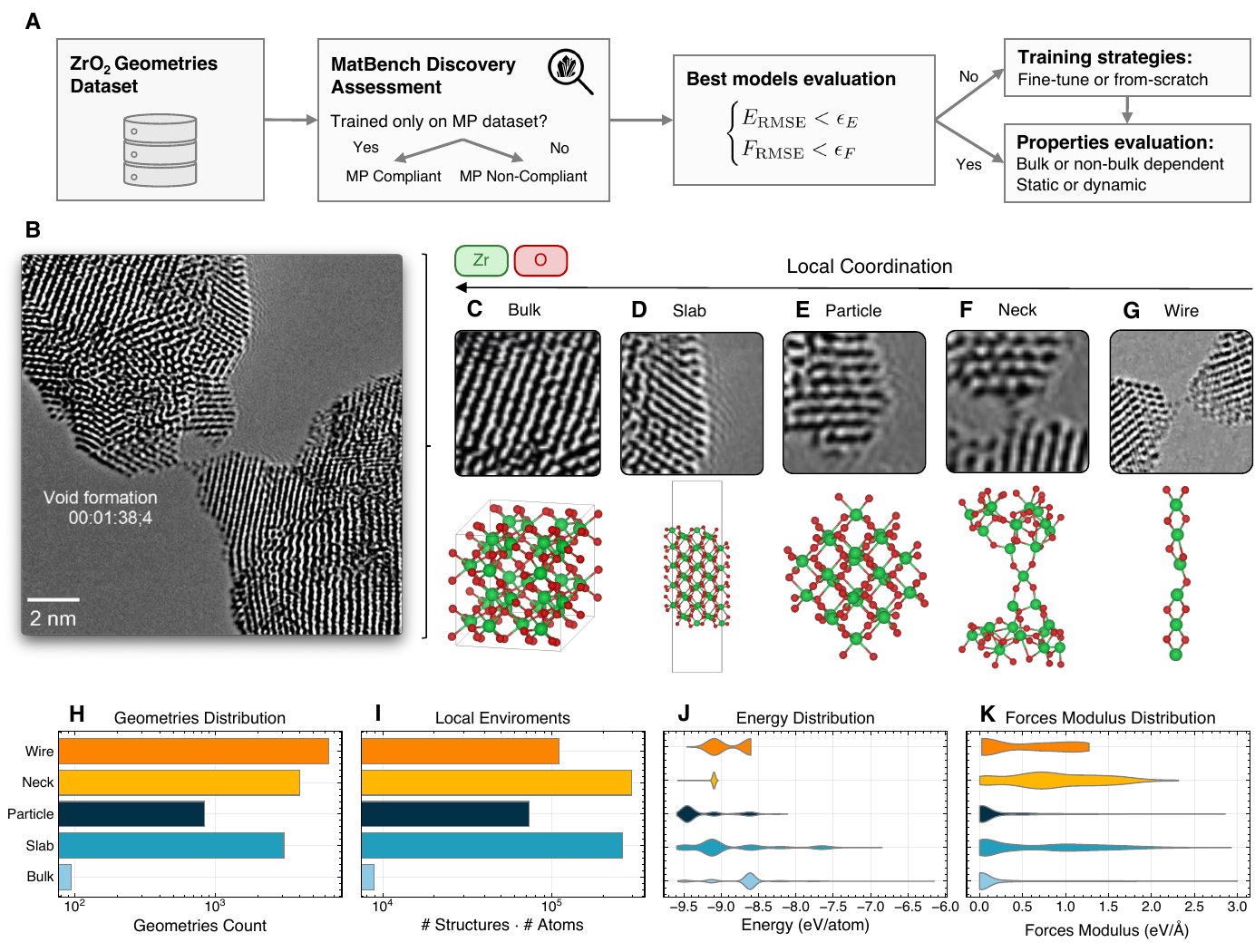}
    \caption{Overview of the ZrO$_2$ cross-geometry dataset and evaluation workflow. (A) Workflow comprising dataset construction, model evaluation and adaptation, and validation using derived physical properties. (B) High-resolution transmission electron microscopy image of a ZrO$_2$ desintering event at the interface between two  \cite{focassio_rev_letter_2022,fiuza_cell_2024}. (C--G) Representative bulk, slab, particle, neck, and wire configurations, respectively, arranged from comparatively high- to low-coordination environments. (H) Number of configurations in each geometry class. (I) Total atomic environments, defined as the sum of the number of atoms over all configurations in each class. (J) Per-atom energy distributions. (K) Atomic-force distributions after applying the 3~eV~\AA$^{-1}$ filtering criterion.}
    \label{fig:geometries}
\end{figure*}

\section{Methods}

\subsection{ZrO$_2$ dataset generation}

The atomic structures used in this work were generated through first-principles simulations based on DFT, as implemented in VASP. The dataset was designed to capture coordination environments arising during the structural evolution of zirconia under tensile deformation, ranging from bulk-like configurations to low-dimensional nanostructures.

To reproduce the experimentally observed desintering process \cite{focassio_rev_letter_2022}, an initial bipyramidal ZrO$_2$ cluster was constructed to represent two connected grains separated by a neck-like region. The structure was subjected to \textit{ab initio} molecular dynamics (AIMD) at 300~K \cite{aimd_2005} while tensile deformation was imposed by progressively separating the outer regions of the cluster. This procedure sampled non-equilibrium configurations involving atomic migration, bond rearrangement, coordination reduction, and eventual neck thinning.

Atomically thin ZrO$_2$ chains were also constructed to probe highly under-coordinated environments. Stoichiometric and oxygen-deficient configurations were elongated quasistatically at 0~K by incrementally increasing the simulation-cell length along the wire axis and thermalizing at 300 K after each increment. Nanoparticle configurations were generated as precursors to the neck structures, while monoclinic ZrO$_2$ slabs were included to sample under-coordinated surface environments. The resulting dataset contains five geometry classes: bulk, slab, particle, neck, and wire, as shown in Fig. \ref{fig:geometries}.

\subsection{Geometry-stratified splitting and force filtering}

The structures were first divided within each geometry class into training, validation, and test partitions. The force filter was applied only after this geometry-stratified assignment, and structures were not reassigned between partitions after filtering. Structures containing an atomic force magnitude greater than 3~eV~\AA$^{-1}$ were removed. Because the fraction of removed configurations differed among geometry classes and partitions, the retained dataset contains approximately 72\% training, 18\% validation, and 10\% test structures rather than the nominal pre-filtering proportions. All reported supervised metrics were calculated using the retained partitions.

This procedure preserves the original partition membership and avoids altering the test set in response to model performance. The final split percentages should therefore be interpreted as the consequence of applying the same physical-data filter independently to the predefined partitions.

\subsection{Zero-shot inference and reference-energy alignment}

For the zero-shot benchmark, we used our Python package IP-Orch \cite{zanineli_2026} to orchestrate model loading, dataset evaluation, and metric calculation. Open-weight Materials Project-compliant (MP-C) and non-compliant (MP-NC) models that fit on an NVIDIA L40 GPU were loaded as Atomic Simulation Environment calculators \cite{ase-paper}. Energies and forces were evaluated for every retained test configuration.

Absolute energies predicted by pretrained models may contain systematic offsets arising from differences in elemental reference energies, training datasets, or electronic-structure settings. To separate these offsets from configuration-dependent errors, the retained training partition was used as the reference dataset for energy alignment. For each model, element-specific  \cite{ceder_2025,kim_2026} were determined by minimizing the energy residuals over the training configurations,
\begin{equation}
\left\{ \Delta\mu_{\alpha} \right\} = \operatorname*{arg\,min}_{\left\{ \Delta\mu_{\alpha} \right\}}
\sum_{i\in\mathrm{train}}
\left[
E_i^{\mathrm{DFT}} - E_i^{\mathrm{model}} - \sum_{\alpha} N_{i\alpha}\Delta\mu_{\alpha}
\right]^2,
\end{equation}
where $N_{i\alpha}$ is the number of atoms of element $\alpha$ in configuration $i$, and $\Delta\mu_{\alpha}$ is the corresponding fitted reference-energy correction. The aligned energy of configuration $i$ was then calculated as
\begin{equation}
E_i^{\mathrm{aligned}} = E_i^{\mathrm{model}} + \sum_{\alpha} N_{i\alpha}\Delta\mu_{\alpha}.
\end{equation}
The corrections obtained from the training partition were held fixed and applied without refitting to the validation and test partitions. Therefore, no validation or test energies were used to determine the alignment parameters. Because the correction depends only on elemental composition, it does not affect the predicted forces.

The aligned energy RMSE is used for the main cross-model comparison. For transparency, the raw, unaligned energy RMSE, the fitted $\Delta\mu_{\alpha}$ values, and relative-energy errors along fixed-composition trajectories are also reported in the Supplementary Information. These complementary quantities distinguish systematic reference-energy offsets from errors in the configuration-dependent shape of the learned potential-energy surface.

Further reference-energy alignment diagnostics, including raw and aligned zero-shot errors, fitted elemental offsets, and relative-energy sequences at fixed composition, are provided in Sec.~II of the Supporting Information (SI), Figs.~S1--S2, and Table~S3.

\subsection{Training protocols}

Training and inference for MACE and ORB were performed using the Graph-PES framework \cite{graphpes_2024}. Unless otherwise stated, fine-tuning refers here to full-parameter, single-head fine-tuning from the pretrained checkpoint, rather than parameter-efficient or replay-based variants such as LoRA, layer freezing, or multihead replay \cite{tompa_2026}. The selected checkpoints were MACE-MP Large and ORB-V3 Direct-20 OMAT. In the comparison, both fine-tuned and from-scratch models were optimized for 50 epochs using the same retained data partitions. The loss included per-atom energy and force terms, and optimization used Adam with a learning rate of $5\times10^{-3}$. Reference energies were shifted during training according to the same training-only convention used for evaluation. This choice follows recent evidence that reference-energy consistency is a prerequisite for stable MLIP fine-tuning and can affect performance as strongly as the fine-tuning strategy itself \cite{ceder_2025,tompa_2026}.

The 50-epoch protocol is designed as a controlled optimization comparison rather than a converged-training comparison. Fine-tuning and from-scratch training required similar wall-clock times under this protocol. Consequently, lower fine-tuning errors indicate a benefit from pretrained initialization at a fixed number of optimization epochs; they do not establish that fine-tuning is intrinsically less computationally expensive.

Complete checkpoint identifiers, optimization settings, and measured wall-clock times are reported in Sec.~I and Table~S1 of the Supporting Information.

\subsection{Geometry-specific and mixed-geometry sampling}

The data is expressed as the \emph{total atomic environments}, defined as the sum of the number of atoms across all configurations in a subset,
\begin{equation}
N_{\mathrm{env}} = \sum_{i=1}^{N_{\mathrm{structures}}} N_i.
\end{equation}
For fixed-size configurations, this quantity reduces to the product of the number of structures and the number of atoms per structure. The term measures the number of atomic environments presented during training, but it does not imply that all environments are structurally independent.

For geometry-specific fine-tuning, each geometry class was divided into four nested subsets with increasing total atomic environments. Each resulting model was evaluated both on its corresponding geometry and on every other geometry class. For mixed-geometry fine-tuning, configurations from bulk, slab, particle, neck, and wire classes were combined while controlling the total atomic environments. This comparison tests whether geometric coverage improves transfer relative to adaptation on an individual structural class.

The geometry-resolved composition and atomic environments of the final filtered training partition are provided in Table~S2 of the SI.

\subsection{Elastic and surface properties}

Elastic properties were calculated using MatCalc \cite{liu2024matcalc}, following the standard energy--strain approach for first-principles mechanical-property calculations \cite{abinitio_2015}. Each model was used to optimize monoclinic ZrO$_2$ with BFGS until the maximum force was below $1\times10^{-6}$~eV~\AA$^{-1}$. Normal and shear strains in the interval $\pm0.004$ were then applied. The energy response was used to obtain the Voigt--Reuss--Hill bulk modulus $K$ and shear modulus $G$. Values in eV~\AA$^{-3}$ were converted to GPa using a factor of 160.2177, and Young's modulus $E$ and Poisson's ratio $\nu$ were derived from $K$ and $G$.

Surface energies were computed from total energies of bulk and symmetric slab configurations. The surface area was calculated as $A=\lVert\mathbf{a}\times\mathbf{b}\rVert$, and
\begin{equation}
\gamma = 16.02\,\frac{E_{\mathrm{slab}}-N E_{\mathrm{bulk}}}{2A},
\end{equation}
where $N=N_{\mathrm{slab}}/N_{\mathrm{bulk}}$, 16.02 converts eV~\AA$^{-2}$ to J~m$^{-2}$, and the factor of two accounts for the two surfaces of a symmetric slab.

Some evaluated surface orientations or closely related slab configurations were represented in the training data. The surface-energy analysis is therefore not treated as a strict out-of-distribution test. Instead, it evaluates whether models reproduce a derived physical observable over a partially overlapping surface domain. Generalization to entirely unseen facets requires a separate facet-held-out assessment.

\subsection{Vibrational properties}

Phonon dispersions were calculated for monoclinic ZrO$_2$ using the finite-displacement method implemented in Phonopy through MatCalc \cite{liu2024matcalc}. Second-order force constants were obtained using a $2\times2\times2$ supercell and a displacement amplitude of 0.015~\AA. The dispersions were evaluated along the same Setyawan--Curtarolo path and q-point sampling used for the Materials Project reference structure mp-2858.

Band RMSE and MAE were calculated by sorting the 36 frequencies at each of the 112 q-points before comparison with the corresponding DFT frequencies. The vibrational-DOS curves were normalized to unit integrated area and linearly interpolated onto a common 3000-point grid over their shared frequency interval. Frequency coordinates below 0~THz were retained to preserve imaginary modes. Additional computational details and complete vibrational results are provided in Sec.~IV of the SI.

\subsection{Molecular dynamics under tensile loading and RMSD analysis}

Molecular dynamics simulations were performed using the Atomic Simulation Environment (ASE) \cite{ase-paper}. Initial neck stability was evaluated through \SI{20}{ps} microcanonical trajectories using velocity Verlet with a \SI{1}{fs} timestep. Initial velocities were sampled at \SI{300}{K}, and configurations were stored every 10 steps. Energy conservation was monitored using
\begin{equation}
\Delta E(t)=E(t)-E(0).
\end{equation}

The mechanical stability of the neck was then evaluated by progressively pulling the structure along the tensile direction. Simulations started from the same DFT configuration and followed the loading protocol of the DFT AIMD reference \cite{focassio_rev_letter_2022}, without additional optimization or atomic constraints. The trajectories were propagated in the NVT ensemble at \SI{300}{K} using a Nosé--Hoover chain thermostat, a damping time of \SI{100}{fs}, and a \SI{1}{fs} timestep.

After approximately \SI{2}{ps} of equilibration, the simulation cell was elongated in steps of \SI{0.5}{\angstrom}, with approximately \SI{2}{ps} of dynamics between each step. Atomic coordinates were scaled with the cell, resulting in a total elongation of \SI{10}{\angstrom} over \SI{42.096}{ps}. The simulations were repeated using three different random seeds.

Atomic trajectories were unwrapped across periodic boundaries before analysis. Each frame was aligned with the initial configuration using the Kabsch algorithm, and the RMSD was calculated as
\begin{equation}
\mathrm{RMSD}(t)=\sqrt{\frac{1}{N}\sum_{i=1}^{N}
\left\lVert\mathbf{r}_i^{(\mathrm{aligned})}(t)
-\mathbf{r}_i^{(\mathrm{ref})}\right\rVert^2}.
\end{equation}
The minimum Zr--O distance was defined as the shortest periodic Zr--O separation in each frame. These quantities were used to monitor structural deformation and bond separation as the neck was subjected to tension.

\section{Results and Discussion}

\begin{figure*}[ht!]
    \centering
    \includegraphics[width=1\linewidth]{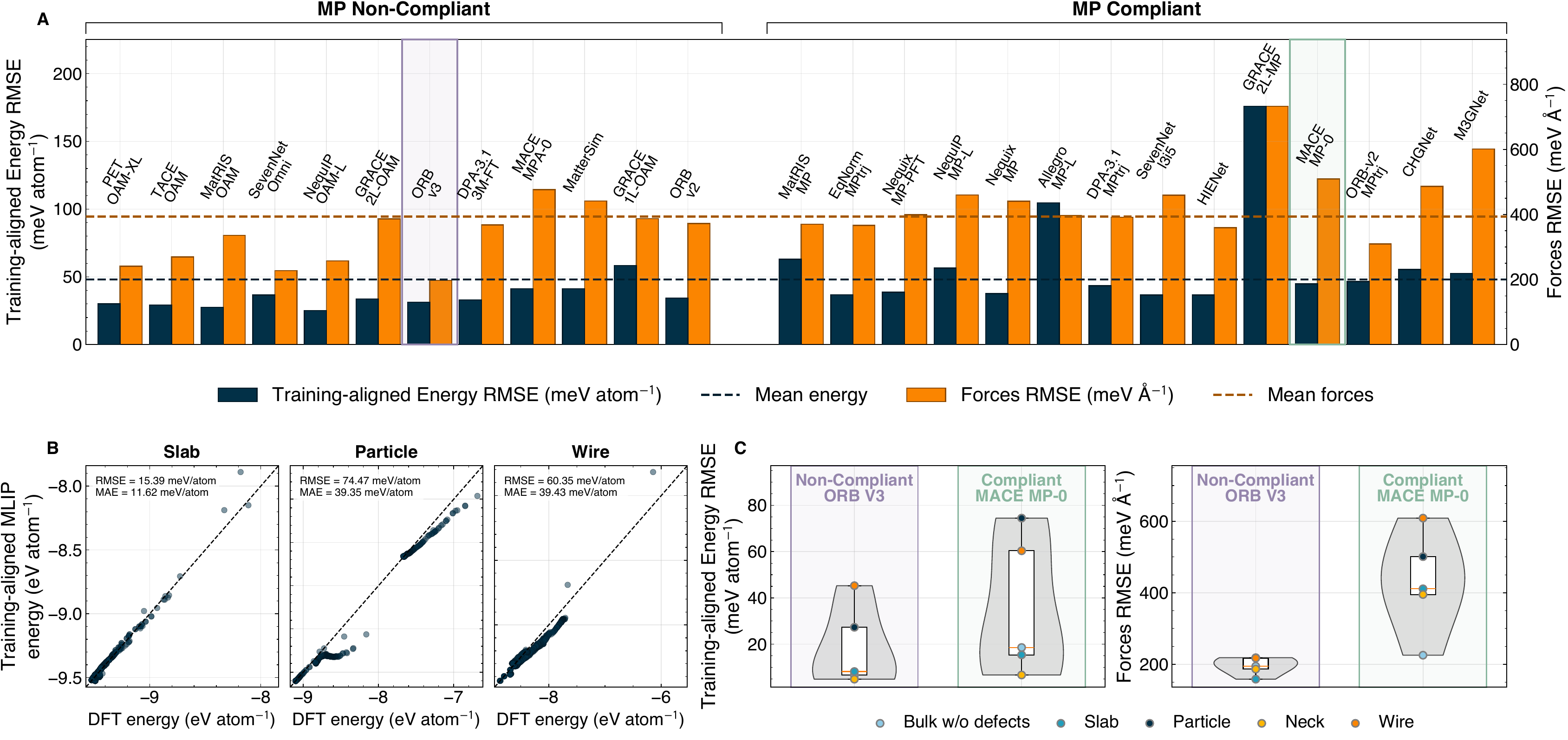}
    \caption{Zero-shot evaluation on the ZrO$_2$ dataset. (A) Aligned per-atom energy RMSE and force RMSE for the evaluated MLIPs, grouped according to the Matbench Discovery MP-NC and MP-C categories. In panel A, black bars indicate aligned per-atom energy RMSE, orange bars indicate force RMSE, and dashed horizontal lines indicate the corresponding mean errors across all evaluated models. (B) Parity plots of training-aligned MACE-MP-0 energies against DFT reference energies for the slab, particle, and wire geometries. (C) Energy and force RMSE distributions for ORB-V3 and MACE-MP-0. Colored markers indicate the errors for the bulk without defects, slab, particle, neck, and wire subsets.}
    \label{fig:zro2_inference}
\end{figure*}

\subsection{Dataset and evaluation framework}

Figure~\ref{fig:geometries} summarizes the experimentally motivated desintering process, the five geometry classes, and the statistical composition of the resulting dataset. The configurations span bulk, slab, particle, neck, and wire environments, progressing from comparatively high coordination in bulk-like structures to strongly under-coordinated wire motifs. The dataset therefore provides a controlled setting for evaluating how pretrained MLIPs respond to changes in coordination and dimensionality within a fixed chemical system.

The distribution of structures alone does not fully represent the amount of information supplied to an atom-centered model because the geometry classes contain different numbers of atoms. We therefore report both the number of configurations and the total atomic environments. Energy and force distributions also vary among geometries, with broader distributions in several low-coordination classes. These differences motivate geometry-resolved evaluation rather than relying only on a single aggregate error.

\subsection{Zero-shot benchmark across zirconia geometries}

Figure~\ref{fig:zro2_inference}A compares the aligned energy and force errors of the evaluated pretrained models. Following the Matbench Discovery convention, the models are grouped as MP-C or MP-NC according to the availability of their code and weights and the data permitted by the benchmark definition \cite{mptrj_2023}. The MP-NC group contains SevenNet MF-ompa \cite{sevenn}, DPA3-v2-OpenLAM \cite{dpa}, GRACE 2L-OAM \cite{grace}, MatterSim \cite{mattersim}, MACE MPA-0 \cite{mace}, GRACE 1L-OAM \cite{grace}, ORB-V2 \cite{orb}, and ORB-V3 \cite{orbv3}. The MP-C group contains SevenNet-i3l5 \cite{sevenn}, ORB-V2-MPtrj \cite{orb}, GRACE 2L-MPtrj \cite{grace}, MACE MP-0 \cite{mace}, and DPA3-v2-MPtrj \cite{dpa}.

Across all models, the mean aligned energy and force RMSEs are around 20~meV atom$^{-1}$ and 400~meV~\AA$^{-1}$, respectively. ORB-V3 is the best-performing MP-NC model, with an energy RMSE of 6~meV atom$^{-1}$ and a force RMSE of 197.3~meV~\AA$^{-1}$. Within the MP-C group, ORB-V2-MPtrj reaches 107.67~meV atom$^{-1}$ and 309.1~meV~\AA$^{-1}$. Models categorized as MP-NC achieve lower errors as a group in this benchmark; however, the comparison does not isolate whether this difference originates from training-set coverage, model architecture, objective functions, or residual reference-energy conventions.

Even the lowest force error remains above commonly used accuracy targets for stable and quantitatively reliable dynamics \cite{grabowski_2023,song_2023}. The zero-shot results therefore motivate target-domain adaptation before these models are applied to the zirconia desintering process. ORB-V3 is selected as the representative MP-NC model for subsequent experiments, while MACE-MP is selected as an MP-C model from a distinct architecture family.

Figures~\ref{fig:zro2_inference}B and \ref{fig:zro2_inference}C resolve the errors by geometry for MACE and ORB, respectively. Bulk configurations are described more accurately than most reduced-coordination configurations. Neck structures show the largest force errors, while wire and slab configurations also produce substantial degradation. These observations are consistent with extrapolation to structural environments that are comparatively rare in bulk-dominated pretraining datasets \cite{gauss_aprox_2010,practical_guida_mlips}. Missing long-range electrostatic physics may also contribute to residual errors in ionic, low-dimensional zirconia. However, the improvement obtained after mixed-geometry fine-tuning indicates that limited target-domain coverage across coordination environments is the dominant practical bottleneck in the present benchmark.

\subsection{Training comparison of zero-shot, fine-tuned, and from-scratch models}

\begin{figure}[ht!]
    \centering
    \includegraphics[width=1\linewidth]{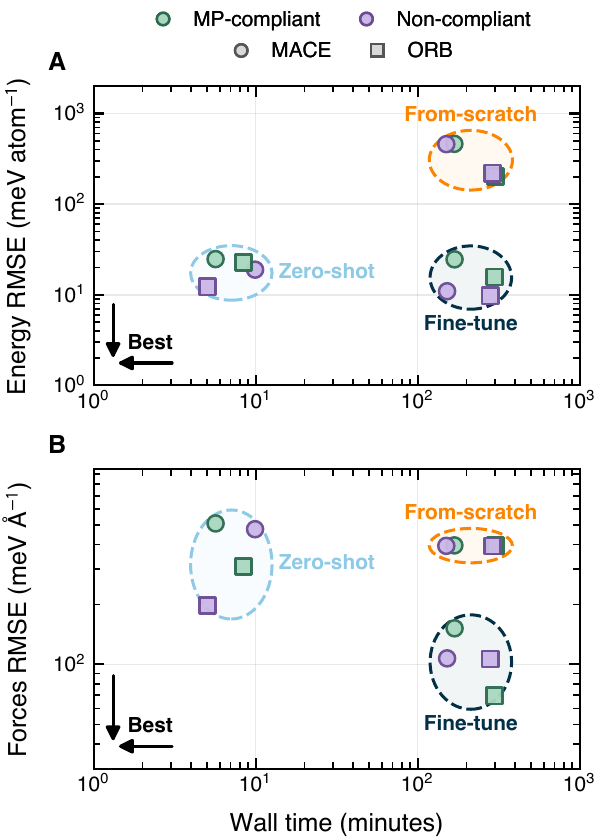}
    \caption{Accuracy–cost comparison of zero-shot, fine-tuned, and from-scratch models. Results are shown for MACE (circles) and ORB (squares), with MP-compliant models in green and MP-non-compliant models in purple. Panels (A) and (B) report the training-aligned energy RMSE and force RMSE, respectively, as functions of wall time. Zero-shot timings correspond only to inference and are included as descriptive baselines rather than cost-equivalent comparisons with the trained models. Lower values along both axes indicate better performance.}
    \label{fig:modalities_comparison}
\end{figure}

Figure~\ref{fig:modalities_comparison} compares zero-shot inference (ZS), fine-tuning (FT), and from-scratch training (FS). For FT and FS, both architectures were trained for the same 50 epochs and evaluated on the same retained test partition. Zero-shot inference provides an immediate baseline but retains the large errors observed for underrepresented geometries. Under the fixed training setup, fine-tuned models achieve lower energy and force errors than their from-scratch counterparts, indicating that pretrained initialization provides a more favorable starting point for adaptation. This result is consistent with recent systematic fine-tuning benchmarks showing that, once reference-energy conventions and stable hyperparameters are controlled, fine-tuned MLIP foundation models often surpass from-scratch models in target-task accuracy, especially under limited-data or limited-optimization regimes \cite{tompa_2026}.

The timing comparison must be interpreted carefully. Fine-tuning and from-scratch training require similar wall-clock times because both optimize the full model for the same number of epochs. Thus, the present result demonstrates improved accuracy at an equal epoch configuration, not a reduction in training cost. Zero-shot timing corresponds only to inference and is therefore not directly equivalent to the one-time optimization cost shown for FT and FS. %

\begin{figure*}[ht!]
    \centering
    \includegraphics[width=1\linewidth]{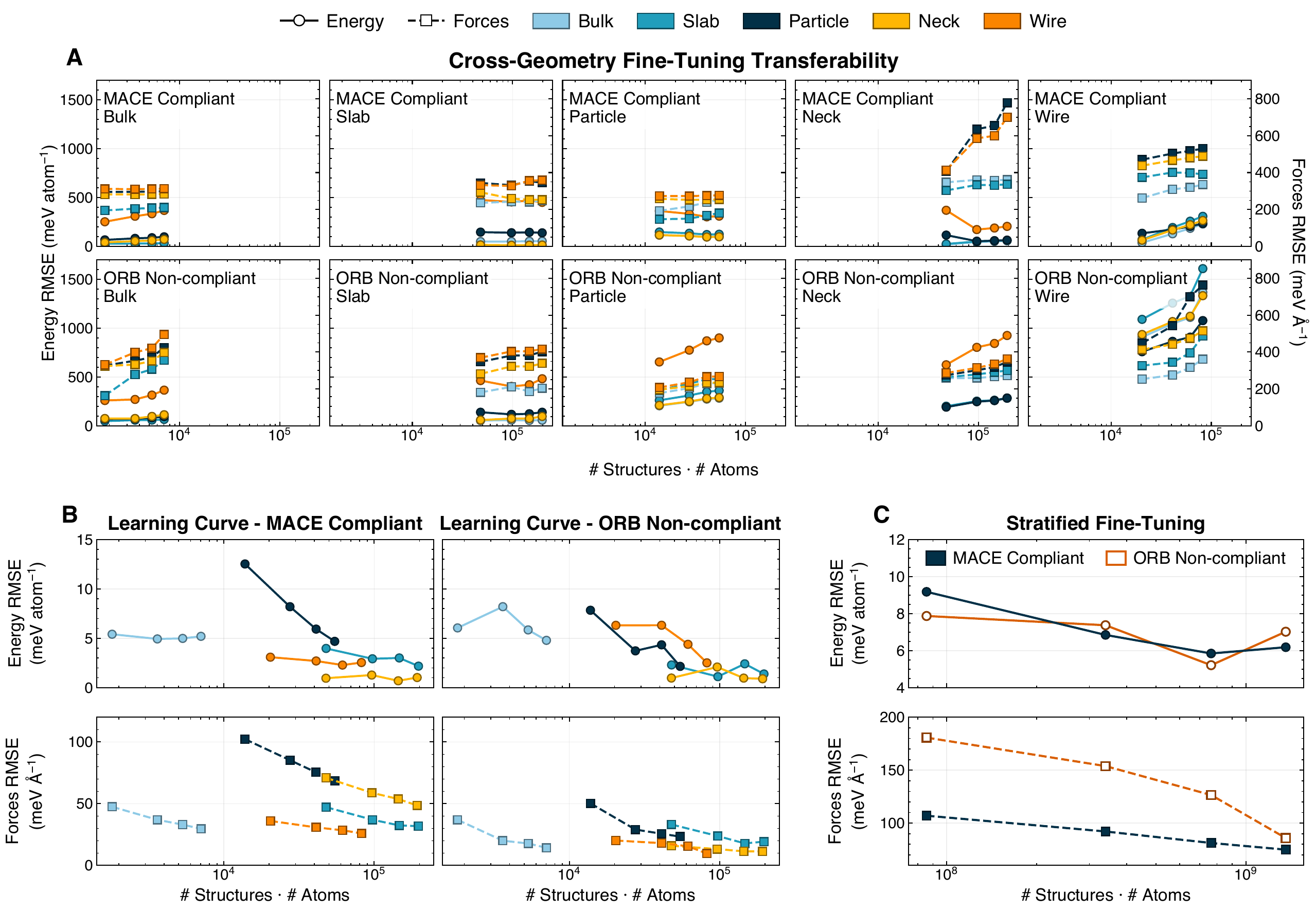}
    \caption{Geometry-specific and mixed-geometry fine-tuning of MACE and ORB on the ZrO$_2$ dataset, assessing in- and out-of-domain geometries. Energy errors are shown with solid lines and circles, and force errors with dashed lines and squares. Colors indicate the evaluation geometry: bulk, slab, particle, neck, and wire. (A) Cross-geometry fine-tuning transferability for MACE (top row) and ORB (bottom row). Each column denotes the geometry used for fine-tuning, and the curves report aligned energy and force RMSE as functions of the total number of atomic environments. (B) Geometry-specific fine-tuning learning curves for MACE and ORB, showing aligned energy and force RMSE as functions of the total number of atomic environments for the bulk, slab, particle, neck, and wire subsets, with evaluation performed on the corresponding geometry. (C) Stratified mixed-geometry fine-tuning as a function of the total number of atomic environments, using training sets composed of configurations drawn from all five geometry classes. This controlled setup isolates the effect of geometric diversity from the effect of simply increasing the amount of training data.}
    \label{fig:full_panel}
\end{figure*}

\subsection{Geometry-specific learning curves}

To determine how the amount and type of target data affect adaptation, each model was fine-tuned using nested subsets from one geometry class at a time. The subset size is reported using the total atomic environment. Figure~\ref{fig:full_panel}A shows the resulting in-geometry learning curves.

For both architectures, increasing the amount of target data generally reduces validation errors, but the rate and stability of improvement depend on geometry. Bulk configurations exhibit comparatively low errors at small data. Slab and particle errors decrease progressively with additional data, whereas neck and wire configurations show larger variability and generally require greater coverage. The result indicates that the number of training configurations alone is insufficient to characterize sample complexity; coordination diversity and the distribution of atomic environments also matter.

Under the checkpoints and training protocol considered here, MACE exhibits more stable cross-geometry force errors than ORB. ORB achieves competitive energy errors for some subsets but shows larger force errors for several low-coordination classes. This comparison is specific to the selected checkpoints, pretraining datasets, and optimization settings and should not be interpreted as a general ranking of the two architecture families.

\subsection{Cross-geometry transfer and mixed-geometry fine-tuning}

The cross-geometry matrices in Figure~\ref{fig:full_panel}B quantify how adaptation on one structural class affects performance on the others. Models fine-tuned only on bulk configurations preserve low bulk errors but deteriorate when evaluated on slabs, necks, and wires. Fine-tuning on slabs or particles improves related environments but does not consistently transfer to the most under-coordinated motifs. Neck and wire configurations remain the most difficult evaluation domains across most training conditions.

Geometry-specific adaptation can also reduce performance outside the target class. In particular, wire-only fine-tuning increases energy errors on several other geometries. This behavior is described here as \emph{negative transfer}: optimizing for a narrow target distribution improves in-domain behavior while degrading performance elsewhere. This distinction is important because full-parameter fine-tuning is expected to be effective when deployment is restricted to the target distribution, but broader deployment requires explicit tests of out-of-distribution robustness and may benefit from replay-based or continual-learning strategies \cite{tompa_2026,reewc_2026}. In this sense, the negative transfer observed here is closely related to catastrophic forgetting: adaptation to a narrow geometry distribution can overwrite parts of the pretrained representation that remain useful for other structural classes.

Under the checkpoints and training protocol considered here, MACE shows more controlled increases in force error across transfer directions than ORB. Nevertheless, both models are sensitive to changes in geometry, demonstrating that adaptation to one class does not ensure reliability across a process that spans multiple coordination regimes.

Figure~\ref{fig:full_panel}C evaluates mixed-geometry fine-tuning at controlled total atomic environments. For MACE, aligned energy and force errors decrease as additional mixed-geometry data are incorporated, with the force RMSE decreasing from approximately 100 to 70~meV~\AA$^{-1}$. ORB shows a similar overall improvement, although its curves are less stable. Relative to the single-geometry models at comparable atomic-environment configurations, exposure to all five classes reduces the disparity among evaluation geometries.

The result supports a data-centric conclusion: target-domain diversity is important when the intended simulation traverses several coordination environments. This conclusion is consistent with recent fine-tuning and transferability benchmarks showing that combined-domain training can improve the uniformity of errors across related structural domains, even though single-domain fine-tuning may remain competitive or superior when deployment is restricted to one domain \cite{tompa_2026,kim_2026,ceder_2025}. It does not establish that stratification solves transferability in general or that data composition is more important than architecture in every setting. Rather, both selected checkpoints benefit from a training set that represents the structural manifold on which they are expected to operate.

\subsection{Property-level validations}

\begin{figure*}[hp!]
    \centering
    \includegraphics[width=.9\linewidth]{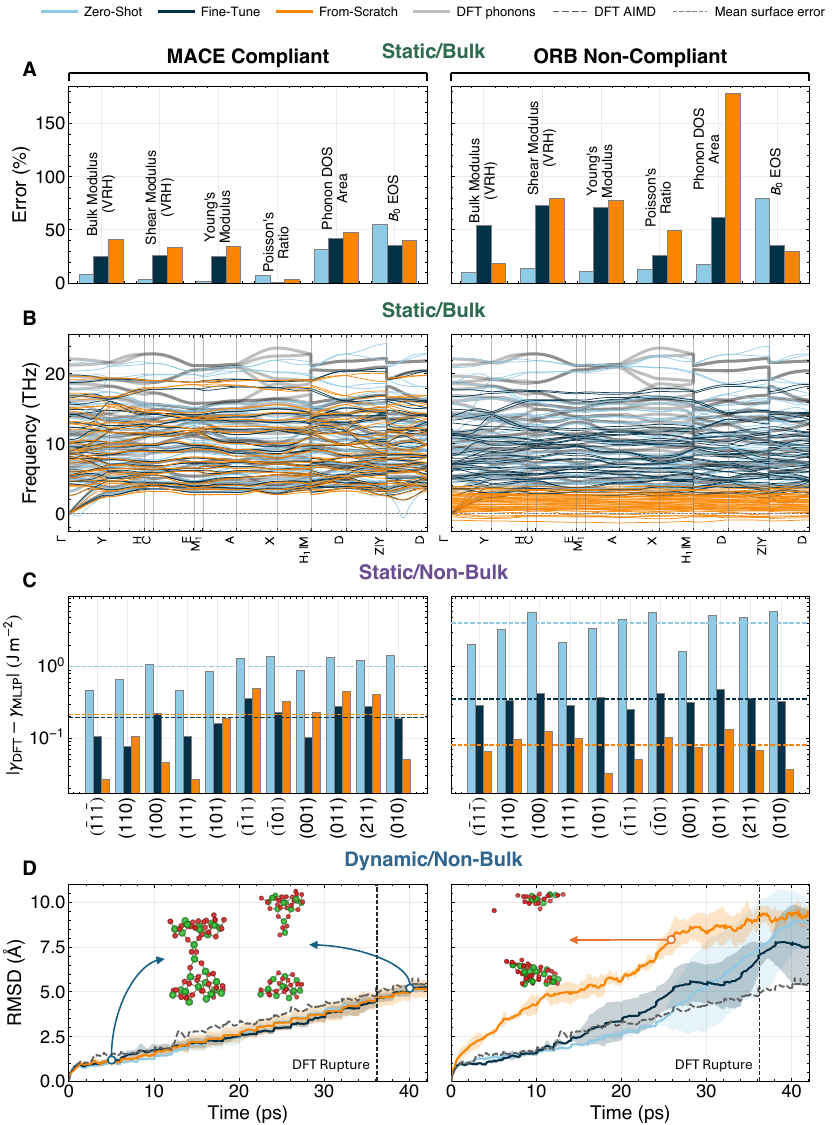}
    \caption{Progressively farther from equilibrium property-level evaluations of zero-shot, fine-tuned, and from-scratch MACE and ORB models. (A) Relative errors in different bulk elastic properties with respect to DFT. (B) Phonon dispersions compared with the DFT reference. (C) Surface-energy errors for the evaluated monoclinic ZrO$_2$ orientations; because some related slab configurations occur in the training data, this panel represents a partially interpolative property test. (D) Neck molecular-dynamics RMSD as a function of time for trajectories, accompanied by representative final configurations.}
    \label{fig:PES_derived}
\end{figure*}

Average energy and force errors do not necessarily determine the accuracy of derived physical quantities. We therefore evaluate the models using a hierarchy of physical tests that progressively move away from near-equilibrium bulk environments. Bulk elastic and vibrational properties probe the response of the crystalline structure near equilibrium; surface energies test static broken-symmetry environments with reduced coordination; and neck molecular dynamics probes a far-from-equilibrium structural evolution in which force errors can accumulate during time integration. This hierarchy, summarized in Figure~\ref{fig:PES_derived} and Table~\ref{tab:mlip_errors}, tests whether improvements in supervised metrics translate into physically reliable behavior across increasing levels of structural complexity.

Related cross-geometry validation has been successful in metallic systems, for example in Gaussian-approximation potentials for Pt trained and tested across bulk, surface, and nanoparticle environments \cite{pt_gap_2023}. The present ZrO$_2$ case is more stringent because dimensionality reduction occurs in a strongly ionic oxide, where low coordination can also modify electrostatic screening and charge redistribution.

\subsubsection{Elastic properties}

For the bulk-dependent properties in Figure~\ref{fig:PES_derived}A, zero-shot models exhibit the lowest errors for most quantities, with the stated MACE exceptions for the bulk modulus and Poisson's ratio. This result is consistent with the strong representation of bulk crystalline environments in the pretraining data. Target-domain adaptation does not uniformly improve bulk elasticity because the fine-tuning dataset also emphasizes surfaces, necks, particles, and wires, while from-scratch training is limited to the available zirconia configurations.

Under the tested settings, MACE provides more accurate elastic predictions than ORB despite its MP-C classification. This observation should be interpreted as checkpoint- and property-specific rather than as a general consequence of compliance status or architecture. Reporting the absolute DFT and predicted values alongside relative errors is necessary to distinguish small percentage differences from physically significant deviations. The corresponding absolute DFT and MLIP values and their percentage deviations are reported in Tables~S4 and S5 of the SI. Comparing both the absolute values and relative deviations shows that fine-tuning generally reduces the discrepancy from DFT, whereas the from-scratch models exhibit larger and more property-dependent errors.

\subsubsection{Vibrational properties}

Figure~\ref{fig:PES_derived}B compares the phonon dispersions predicted by the zero-shot, fine-tuned, and from-scratch models with the DFT reference. The zero-shot models provide the closest overall agreement with the DFT dispersion, with phonon eigenvalue RMSEs of 0.785~THz for MACE and 0.540~THz for ORB. Fine-tuning increases these errors to 1.653 and 3.067~THz, respectively, while the from-scratch models reach 1.598~THz for MACE and 10.446~THz for
ORB.

These results show that improving energy and force errors on the
geometry-diverse zirconia dataset does not necessarily preserve near-equilibrium vibrational behavior. In particular, the strong degradation of the ORB from-scratch dispersion demonstrates that a model can achieve useful configuration-level accuracy while failing to reproduce the curvature of the potential-energy surface around the crystalline reference structure. Phonon calculations therefore provide a complementary validation of model quality that is not captured by aggregate energy and force metrics alone.

Complete phonon dispersions, normalized vibrational-DOS comparisons, and quantitative band- and DOS-error metrics are provided in Figures~S3--S5 and Table~S10 of the SI. The zero-shot comparison between the ORB-v3 Direct-20 and Conservative-20 variants is presented separately in Figure~S4.

\begin{table*}[ht!]
\centering

\caption{Absolute (J~m$^{-2}$) and relative (\%) errors of MLIP
surface energies with respect to DFT. The evaluation is partly
interpolative because some evaluated orientations or closely related
slabs are represented in the training data. Cell colors follow fixed
continuous scales shared by MACE and ORB. Absolute errors are mapped
from 0 to 1.50~J~m$^{-2}$, while relative errors are mapped from
0 to 25\%. Values equal to or above the upper limits are shown with
the maximum red intensity. Lower errors are greener and higher errors
are redder.}

\label{tab:mlip_errors}

\setlength{\tabcolsep}{3pt}
\renewcommand{\arraystretch}{1.12}

\begin{tabularx}{\textwidth}{@{}l*{6}{C}|*{6}{C}@{}}

\toprule

& \multicolumn{6}{c|}{MACE Compliant}
& \multicolumn{6}{c}{ORB Non-Compliant} \\

\cmidrule(lr){2-7}
\cmidrule(lr){8-13}

& \multicolumn{2}{c}{Zero-shot}
& \multicolumn{2}{c}{Fine-tuned}
& \multicolumn{2}{c|}{From scratch}
& \multicolumn{2}{c}{Zero-shot}
& \multicolumn{2}{c}{Fine-tuned}
& \multicolumn{2}{c}{From scratch} \\

\cmidrule(lr){2-3}
\cmidrule(lr){4-5}
\cmidrule(lr){6-7}
\cmidrule(lr){8-9}
\cmidrule(lr){10-11}
\cmidrule(lr){12-13}

Surface
& J~m$^{-2}$ & \%
& J~m$^{-2}$ & \%
& J~m$^{-2}$ & \%
& J~m$^{-2}$ & \%
& J~m$^{-2}$ & \%
& J~m$^{-2}$ & \% \\

\midrule

$(\bar{1}1\bar{1})$
& \heatJ{0.47} & \heatP{11.86}
& \heatJ{0.11} & \heatP{2.71}
& \heatJ{0.03} & \heatP{0.68}
& \heatJ{2.07} & \heatP{52.86}
& \heatJ{0.29} & \heatP{7.47}
& \heatJ{0.07} & \heatP{1.69} \\

$(110)$
& \heatJ{0.66} & \heatP{13.04}
& \heatJ{0.08} & \heatP{1.53}
& \heatJ{0.11} & \heatP{2.09}
& \heatJ{3.38} & \heatP{66.31}
& \heatJ{0.34} & \heatP{6.62}
& \heatJ{0.10} & \heatP{1.90} \\

$(100)$
& \heatJ{1.08} & \heatP{16.20}
& \heatJ{0.23} & \heatP{3.39}
& \heatJ{0.05} & \heatP{0.69}
& \heatJ{5.78} & \heatP{86.56}
& \heatJ{0.43} & \heatP{6.38}
& \heatJ{0.12} & \heatP{1.85} \\

$(111)$
& \heatJ{0.47} & \heatP{11.86}
& \heatJ{0.11} & \heatP{2.71}
& \heatJ{0.03} & \heatP{0.68}
& \heatJ{2.21} & \heatP{56.47}
& \heatJ{0.29} & \heatP{7.46}
& \heatJ{0.10} & \heatP{2.54} \\

$(101)$
& \heatJ{0.87} & \heatP{15.09}
& \heatJ{0.16} & \heatP{2.80}
& \heatJ{0.19} & \heatP{3.32}
& \heatJ{3.46} & \heatP{60.01}
& \heatJ{0.37} & \heatP{6.39}
& \heatJ{0.03} & \heatP{0.57} \\

$(\bar{1}11)$
& \heatJ{1.31} & \heatP{20.06}
& \heatJ{0.36} & \heatP{5.54}
& \heatJ{0.51} & \heatP{7.72}
& \heatJ{4.68} & \heatP{71.42}
& \heatJ{0.25} & \heatP{3.83}
& \heatJ{0.05} & \heatP{0.77} \\

$(\bar{1}01)$
& \heatJ{1.41} & \heatP{19.89}
& \heatJ{0.23} & \heatP{3.20}
& \heatJ{0.33} & \heatP{4.61}
& \heatJ{5.79} & \heatP{81.76}
& \heatJ{0.43} & \heatP{6.08}
& \heatJ{0.10} & \heatP{1.47} \\

$(001)$
& \heatJ{0.90} & \heatP{14.85}
& \heatJ{0.10} & \heatP{1.70}
& \heatJ{0.23} & \heatP{3.83}
& \heatJ{1.65} & \heatP{27.22}
& \heatJ{0.32} & \heatP{5.20}
& \heatJ{0.08} & \heatP{1.23} \\

$(011)$
& \heatJ{1.38} & \heatP{19.64}
& \heatJ{0.28} & \heatP{3.99}
& \heatJ{0.45} & \heatP{6.47}
& \heatJ{5.23} & \heatP{74.61}
& \heatJ{0.48} & \heatP{6.85}
& \heatJ{0.13} & \heatP{1.88} \\

$(211)$
& \heatJ{1.22} & \heatP{17.48}
& \heatJ{0.28} & \heatP{4.01}
& \heatJ{0.41} & \heatP{5.91}
& \heatJ{4.92} & \heatP{70.68}
& \heatJ{0.36} & \heatP{5.20}
& \heatJ{0.07} & \heatP{0.97} \\

$(010)$
& \heatJ{1.45} & \heatP{17.58}
& \heatJ{0.19} & \heatP{2.30}
& \heatJ{0.05} & \heatP{0.61}
& \heatJ{6.01} & \heatP{73.14}
& \heatJ{0.33} & \heatP{4.04}
& \heatJ{0.04} & \heatP{0.45} \\

\bottomrule

\end{tabularx}
\end{table*}%

\subsubsection{Surface energies}

Table~\ref{tab:mlip_errors} shows that the ranking of adaptation strategies depends on the model family. For MACE, fine-tuning gives the lowest mean relative surface-energy error, approximately 3.08\%, compared with 3.33\% for training from scratch and 16.14\% for zero-shot inference. Fine-tuning outperforms the from-scratch model for 7 of the 11 listed surfaces. For ORB, training from scratch is clearly best, with a mean relative error of approximately 1.39\%, compared with 5.96\% for fine-tuning and 65.55\% for zero-shot inference.

The surface-energy evaluation includes orientations or related slab environments represented in the training data and is therefore partly interpolative. The results demonstrate whether low configuration-level errors are sufficient to reproduce a derived surface observable within this partially overlapping domain; they do not by themselves establish generalization to unseen facets. Within this scope, adaptation substantially improves both architectures over their zero-shot baselines, but the optimal strategy differs between MACE and ORB.

\subsubsection{Neck molecular dynamics}

Figure~\ref{fig:PES_derived}D provides an initial dynamical assessment using 20-ps neck trajectories. RMSD is used to identify structural drift relative to the initial configuration after removing rigid translation and rotation. The fine-tuned trajectories remain bounded over most of the simulated interval, while the zero-shot MACE trajectory exhibits a small structural rearrangement. The from-scratch ORB trajectory develops a larger deviation and ruptures under the tested initial condition.

These individual trajectories reveal failure modes that are not apparent from aggregate static metrics, but they should not be interpreted as a statistical proof of dynamical stability. RMSD can reflect either physically meaningful relaxation or model failure, and the current comparison uses a single initial-velocity realization. Three different seeds were used for running the trajectories. Additional trajectory diagnostics, including energy-drift slopes, minimum Zr--O distances, representative snapshots, and pair-distance distributions, are provided in Figs.~S6--S8 and Table~S11 of the SI.

\section{Conclusion}

We evaluated the transferability of foundation MLIPs across a geometry-diverse ZrO$_2$ dataset containing bulk, slab, particle, neck, and wire configurations. The zero-shot benchmark reveals strongly geometry-dependent errors: bulk-like environments are described more accurately, while neck and wire configurations remain challenging. Although MP-NC models achieve lower errors overall in the present benchmark, the available comparison does not isolate the respective effects of pretraining data, architecture, training objectives, and reference-energy conventions.

Under an equal 50-epoch optimization setup, fine-tuning produces lower energy and force errors than training from scratch, while requiring comparable wall-clock time. Geometry-specific fine-tuning improves accuracy within the target class but often produces negative transfer to other classes. In contrast, mixed-geometry fine-tuning at controlled total atomic-environment configurations reduces cross-geometry variability for both tested checkpoints. This result shows that explicit coverage of the structural environments encountered during a simulation is important when adapting pretrained MLIPs to geometry-changing processes. More broadly, it reinforces the idea that the appropriate fine-tuning protocol depends on deployment scope: narrow full-parameter fine-tuning can be effective for a restricted target geometry, whereas simulations that traverse multiple coordination regimes require either geometry-diverse target data or adaptation strategies designed to preserve broader robustness \cite{tompa_2026,kim_2026,ceder_2025}.

Property-level tests demonstrate that no single training strategy is uniformly optimal and that model rankings depend on the physical complexity of the validation target. Bulk elasticity and lattice dynamics, surface energetics, and neck dynamics probe progressively different regions of the potential-energy surface. Zero-shot models retain strong performance for several elastic and vibrational quantities, whereas adaptation provides larger improvements for the partially overlapping surface domain. In the case of monoclinic zirconia, preliminary neck trajectories further expose dynamical behavior not captured by aggregate static errors. Because the surface test is partly interpolative and the MD analysis is based on a single trajectory per model, these results are best interpreted as complementary diagnostics rather than definitive out-of-distribution or stability benchmarks.

In summary, geometry-diverse data and property-specific validation substantially improve the defensibility of MLIP deployment for zirconia nanostructures. In this context, we propose that such a dataset can serve as a benchmark for out-of-distribution structures, complementing existing datasets by incorporating configurations beyond bulk-like environments, such as low-coordination and non-equilibrium geometries \cite{szilvasi_2025}.
The present results suggest that geometry-diverse target data can recover much of the practical accuracy needed to stabilize the tested low-dimensional ZrO$_2$ trajectories, even without an explicit treatment of long-range electrostatics. Remaining questions include whether long-range electrostatic terms are required to reduce the residual errors further, as well as the effects of force-filtering choices, trajectory-level data correlation, and fully converged optimization. Future work should also replace manually defined geometry subsets with automated active-learning workflows that explore the potential-energy surface more systematically. Frameworks such as autoplex illustrate how iterative structure generation, model evaluation, and reference-data acquisition can be combined to reduce manual dataset curation and improve coverage of diverse atomic environments \cite{autoplex_2025}. Addressing these controls will determine whether the remaining low-dimensional errors arise primarily from data coverage or from limitations of the underlying short-range model representations.

\section*{Acknowledgements}
The authors thank the support of São Paulo Research Foundation (FAPESP) for the financial support under grants nº 2024/22392-2, 2024/00989-7, and 2023/13081-0,
from CNPq project nos. 422069/2023-0, 313301/2025-5, and CNPq - INCT (National Institute of Science and Technology on Materials Informatics) grant no. 371610/2023-0.
\subsection*{Data and code availability}
The datasets and code are available at the repository: \url{https://doi.org/10.5281/zenodo.21829037}.

%

\onecolumngrid\clearpage


\section*{Supplementary material}

# Supporting Information
# Cross-Geometry Transferability Assessment of Universal Machine Learning Interatomic Potentials: From bulk materials to atomic nanowires

Pedro H. M. Zanineli [1,2] Bruno Focassio [1] and Gabriel R. Schleder [1,2,*]

[1] *Brazilian Nanotechnology National Laboratory (LNNano/CNPEM), 13083-100, Campinas, SP, Brazil*
[2] *Universidade Federal do ABC (UFABC), 09210-580, Santo André, São Paulo, Brazil*



This Supplementary Information provides details supporting the main text of "Cross-Geometry Transferability Assessment of Foundation Machine Learning Interatomic Potentials: From bulk materials to atomic nanowires".

## CONTENTS



* Corresponding author. gabriel.schleder@lnnano.cnpem.br

# I. MODEL AND TRAINING DETAILS

The model training contain fine-tuning and from-scratch configurations for MACE MP-0, MACE MPA-0, ORB-v2-MPtraj, and ORB-v3 [1–3], trained using GraphPES [4]. All models were trained for 50 epochs using float32 precision, the AdamW optimizer with an initial learning rate of 0.005, a ReduceLROnPlateau scheduler, a batch size of 8, random seed 42, and one NVIDIA L40 GPU. The objective combined per-atom energy loss and force RMSE with implicit unit weights. Fine-tuned models were initialized from their corresponding pretrained checkpoints, whereas from-scratch models retained the same architectures with randomly initialized weights.

These trained models are divided into two categories according to the datasets used during their initial training. Models trained exclusively on the Materials Project dataset [5] are classified as Materials Project Compliant (MP-C), whereas models trained on broader datasets are classified as Materials Project Non-Compliant (MP-NC).

TABLE S1. Model checkpoints and wall times for the training runs. Fine-tuning (FT) starts from the corresponding pretrained checkpoint, whereas from-scratch training (FS) uses the same architecture with randomly initialized weights.

| Model | Category | Checkpoint | FT time (min) | FS time (min) |
|---|---|---|---|---|
| MACE MP-0 | MP-C | large | 169.0 | 168.5 |
| ORB-v2-MPtrj | MP-C | orb-mptraj-only-v2 | 298.6 | 302.2 |
| MACE MPA-0 | MP-NC | medium-mpa-0 | 152.0 | 150.3 |
| ORB-v3 | MP-NC | orb-v3-direct-20-omat | 280.8 | 289.2 |

For a collection of structures indexed by $i$, the atomic-environment is

$$N_{\text{env}} = \sum_i N_i, \tag{S1}$$

where $N_i$ is the number of atoms in structure $i$. Table S2 reports the resulting geometry-resolved for the final training partition.

TABLE S2. Geometry-resolved composition of the final filtered training partition. The atomic-environment, $N_{\text{env}}$, is the total number of atoms across all structures in each geometry class.

| Training geometry | Structures | $N_{\text{env}}$ |
|---|---|---|
| Bulk | 67 | 6 366 |
| Slab/surface defect/GB | 2140 | 153 112 |
| Particle | 548 | 53 911 |
| Neck | 2269 | 167 906 |
| Wire | 4307 | 73 556 |
| Total | 9331 | 454 851 |

## II. REFERENCE-ENERGY ALIGNMENT DIAGNOSTICS

The intended elemental reference-energy correction is

$$E_i^{\text{aligned}} = E_i^{\text{model}} + \sum_{\alpha} N_{i\alpha} \Delta\mu_\alpha, \tag{S2}$$

where $N_{i\alpha}$ counts atoms of element $\alpha$ in configuration $i$. A valid diagnostic requires fitting $\Delta\mu_\alpha$ on the training partition only and then applying the fixed correction to evaluation data. For each model, $\Delta\mu_{\text{Zr}}$ and $\Delta\mu_{\text{O}}$ were fitted by least squares on these 9331 training structures only. The fixed corrections were then applied to the complete evaluation set.

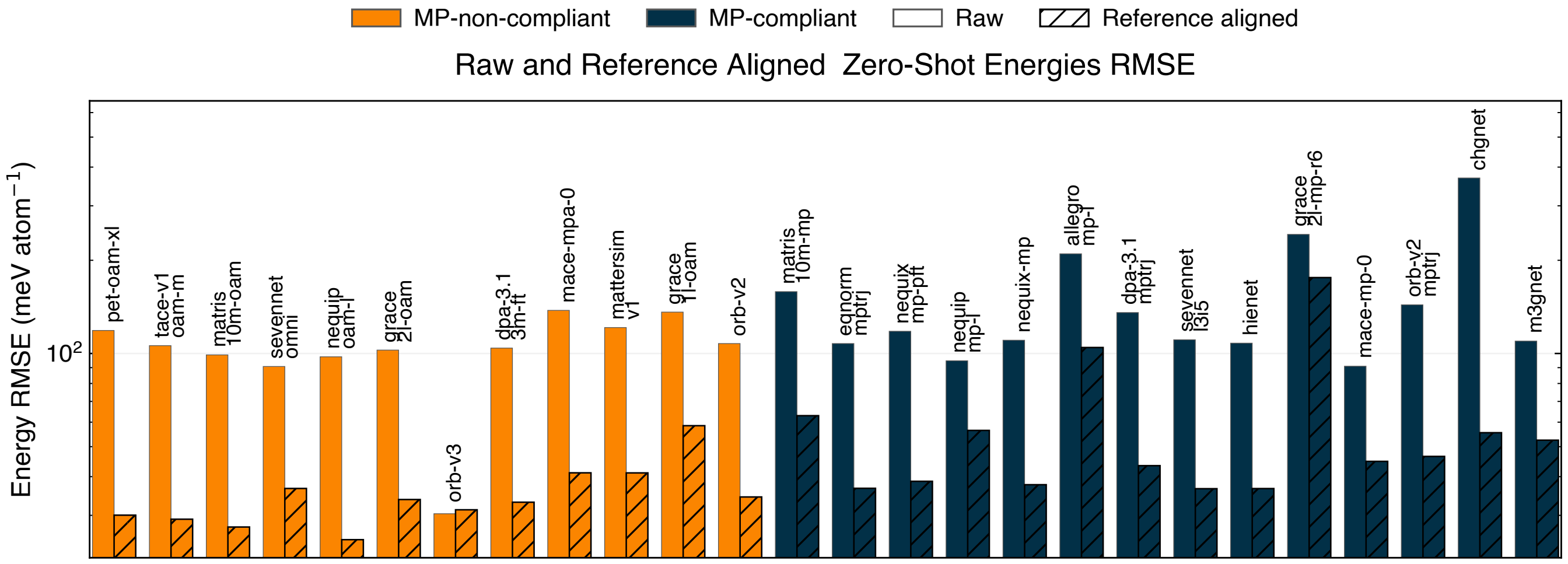


FIG. S1. Raw and training-only elemental-aligned zero-shot energy RMSE in the Matbench Discovery order. Orange and dark-blue fills denote MP-non-compliant and MP-compliant groups, respectively; hatching denotes aligned values.

TABLE S3. Zero-shot energy and force diagnostics. Elemental corrections were fitted by least squares on the 9331 structures in the supplied filtered training split and then held fixed for the complete evaluation set. Energy quantities are in meV atom$^{-1}$, force RMSE is in meV Å$^{-1}$, and $\Delta\mu_\alpha$ is in eV. Cell colors use fixed pastel scales shared by MP-compliant and MP-non-compliant models for each metric: 0–200 meV atom$^{-1}$ for raw energy RMSE and MAE, 0–100 meV atom$^{-1}$ for aligned energy RMSE and MAE, and 0–600 meV Å$^{-1}$ for force RMSE. Values equal to or above an upper limit receive the maximum red intensity. The fitted elemental corrections are not color-coded.

| Model | Category | Raw E RMSE | Aligned E RMSE | Raw E MAE | Aligned E MAE | F RMSE | $\Delta\mu_{Zr}$ | $\Delta\mu_O$ |
|---|---|---|---|---|---|---|---|---|
| | | (meV atom$^{-1}$ or meV Å$^{-1}$) | | | | | (eV) | |
| pet-oam-xl | MP-non-compliant | 118.68 | 30.04 | 115.26 | 21.65 | 240.61 | -0.307129 | -0.001490 |
| tace-v1-oam-m | MP-non-compliant | 106.01 | 29.15 | 102.87 | 14.37 | 269.62 | -0.462935 | 0.074944 |
| matris-10m-oam | MP-non-compliant | 98.96 | 27.51 | 96.35 | 11.91 | 335.61 | -0.450614 | 0.077864 |
| sevennet-omni | MP-non-compliant | 90.78 | 36.64 | 87.74 | 18.77 | 227.62 | -0.330242 | 0.017223 |
| nequip-oam-l | MP-non-compliant | 97.56 | 25.04 | 95.52 | 11.37 | 256.93 | -0.446863 | 0.074769 |
| grace-2l-oam | MP-non-compliant | 102.56 | 33.75 | 99.16 | 20.32 | 386.19 | -0.319289 | 0.000208 |
| orb-v3 | MP-non-compliant | 30.36 | 31.27 | 17.38 | 13.99 | 197.29 | -0.192228 | 0.086643 |
| dpa-3.1-3m-ft | MP-non-compliant | 104.10 | 33.08 | 98.52 | 24.93 | 367.85 | -0.360207 | 0.047175 |
| mace-mpa-0 | MP-non-compliant | 137.80 | 41.16 | 132.44 | 30.63 | 476.44 | -0.379649 | 0.024794 |
| mattersim-v1 | MP-non-compliant | 121.28 | 41.14 | 117.07 | 23.59 | 441.60 | -0.472044 | 0.071268 |
| grace-1l-oam | MP-non-compliant | 136.05 | 58.44 | 126.44 | 48.50 | 387.17 | -0.474536 | 0.030519 |
| orb-v2 | MP-non-compliant | 107.67 | 34.40 | 103.38 | 17.86 | 371.92 | -0.496684 | 0.092579 |
| matris-10m-mp | MP-compliant | 158.41 | 62.97 | 147.65 | 45.88 | 370.35 | -0.505167 | 0.089979 |
| eqnorm-mptrj | MP-compliant | 107.67 | 36.69 | 102.82 | 21.25 | 366.34 | -0.297799 | -0.004384 |
| nequix-mp-pft | MP-compliant | 117.97 | 38.65 | 112.29 | 25.46 | 398.98 | -0.369085 | 0.031462 |
| nequip-mp-l | MP-compliant | 94.72 | 56.43 | 91.46 | 28.18 | 460.01 | -0.444325 | 0.074786 |
| nequix-mp | MP-compliant | 110.36 | 37.68 | 104.92 | 24.22 | 440.75 | -0.314798 | 0.012268 |
| allegro-mp-l | MP-compliant | 209.89 | 104.55 | 190.26 | 78.12 | 397.10 | -0.516534 | 0.069808 |
| dpa-3.1-mptrj | MP-compliant | 135.57 | 43.42 | 129.28 | 32.78 | 391.78 | -0.419245 | 0.053882 |
| sevennet-l3i5 | MP-compliant | 110.73 | 36.58 | 106.86 | 20.43 | 459.17 | -0.442815 | 0.066210 |
| hienet | MP-compliant | 107.98 | 36.62 | 105.06 | 18.47 | 359.52 | -0.352175 | 0.025746 |
| grace-2l-mp-r6 | MP-compliant | 242.48 | 175.87 | 213.17 | 135.36 | 732.26 | 1.913479 | -0.499138 |
| mace-mp-0 | MP-compliant | 90.96 | 44.83 | 87.74 | 23.98 | 509.17 | -0.409972 | 0.060916 |
| orb-v2-mptrj | MP-compliant | 143.61 | 46.51 | 138.40 | 33.54 | 309.14 | -0.531955 | 0.091778 |
| chgnet | MP-compliant | 368.92 | 55.48 | 363.00 | 41.90 | 486.37 | 2.131660 | -0.516475 |
| m3gnet | MP-compliant | 109.65 | 52.47 | 99.79 | 37.89 | 601.14 | -0.176665 | -0.028656 |

## III. COMPLETE PROPERTY VALUES

Table S4 collects the absolute elastic and equation-of-state quantities. Table S5 reports percentual deviations from the mp-2858 DFT reference elastic row [5, 6].

TABLE S4. Elastic, equation-of-state, and phonon-DOS quantities. For the machine-learning models, the reported volume is the equilibrium volume $V_0$ obtained from the equation-of-state fit.

| Model | Volume (Å$^3$) | $K$ (GPa) | $G$ (GPa) | $E$ (GPa) | $\nu$ | $B_0'$ |
|---|---|---|---|---|---|---|
| DFT | 144.467 | 183.000 | 86.000 | 223.058 | 0.297 | 4.257 |
| MACE zero-shot | 144.308 | 168.554 | 89.198 | 227.470 | 0.275 | 3.289 |
| MACE fine-tuned | 144.891 | 228.303 | 108.079 | 280.045 | 0.296 | 7.386 |
| MACE from-scratch | 146.566 | 258.249 | 114.567 | 299.424 | 0.307 | 6.619 |
| ORB zero-shot | 142.819 | 232.313 | 114.050 | 294.033 | 0.289 | -44.337 |
| ORB fine-tuned | 142.995 | 84.474 | 23.263 | 63.920 | 0.374 | 9.116 |
| ORB from-scratch | 144.768 | 149.595 | 17.529 | 50.610 | 0.444 | -4.904 |

TABLE S5. Absolute percentage errors relative to the DFT/reference values in Table S4. Cell colors use a fixed pastel scale shared by MACE and ORB and by all reported properties. Errors are mapped from 0 to 25%; values equal to or above 25% receive the maximum red intensity. Lower errors are greener and higher errors are redder.

| Model | $V_0$ (Å$^3$) | $K$ (GPa) | $G$ (GPa) | $E$ (GPa) | $\nu$ |
|---|---|---|---|---|---|
| MACE zero-shot | 1.864 | 7.894 | 3.719 | 1.978 | 7.343 |
| MACE fine-tuned | 2.275 | 24.756 | 25.673 | 25.548 | 0.438 |
| MACE from scratch | 3.457 | 41.120 | 33.218 | 34.236 | 3.334 |
| ORB zero-shot | 0.812 | 26.947 | 32.616 | 31.819 | 2.627 |
| ORB fine-tuned | 0.936 | 53.839 | 72.950 | 71.344 | 25.935 |
| ORB from scratch | 2.188 | 18.254 | 79.618 | 77.311 | 49.411 |

The complete surface-energy predictions are reported in Table S6 in the exact Miller-index ordering. Table S7 provide the corresponding absolute and relative errors.

TABLE S6. Monoclinic $ZrO_2$ surface energies (J m$^{-2}$).

| Miller index | DFT | mace-zero-shot | mace-fine-tune | mace-from-scratch | orb-zero-shot | orb-fine-tune | orb-from-scratch |
|---|---|---|---|---|---|---|---|
| $(\bar{1}1\bar{1})$ | 3.920 | 3.456 | 3.814 | 3.947 | 1.848 | 4.213 | 3.854 |
| (110) | 5.094 | 4.430 | 5.016 | 5.200 | 1.716 | 5.431 | 4.997 |
| (100) | 6.676 | 5.594 | 6.450 | 6.722 | 0.897 | 7.102 | 6.553 |
| (111) | 3.920 | 3.456 | 3.814 | 3.947 | 1.707 | 4.213 | 3.821 |
| (101) | 5.762 | 4.893 | 5.601 | 5.571 | 2.305 | 6.130 | 5.795 |
| $(\bar{1}11)$ | 6.548 | 5.234 | 6.185 | 6.042 | 1.871 | 6.798 | 6.598 |
| $(\bar{1}01)$ | 7.087 | 5.677 | 6.860 | 6.760 | 1.292 | 7.517 | 7.190 |
| (001) | 6.058 | 5.158 | 5.955 | 6.290 | 4.409 | 6.373 | 5.983 |
| (011) | 7.004 | 5.629 | 6.725 | 6.551 | 1.779 | 7.484 | 6.872 |
| (211) | 6.963 | 5.746 | 6.684 | 6.551 | 2.042 | 7.325 | 6.895 |
| (010) | 8.223 | 6.777 | 8.034 | 8.173 | 2.209 | 8.555 | 8.260 |

Table S8 provides slabs atom count, projected slab thickness, vacuum thickness, and surface area, and exact training overlap.

Exact matches with frames in the filtered training partition were identified for most surface orientations, as summarized in Table **??**. Consequently, the corresponding surface-energy predictions should not be interpreted as strict no-overlap transfer tests. The (111) orientation is the only evaluated surface for which no exact training-frame match

TABLE S7. Absolute (J $m^{-2}$) and relative (%) errors of MLIP surface energies with respect to DFT. The evaluation is partly interpolative because some evaluated orientations or closely related slabs are represented in the training data. Cell colors use fixed pastel scales shared by MACE and ORB. Absolute errors are mapped from 0 to 1.50 J $m^{-2}$, while relative errors are mapped from 0 to 25%. Values equal to or above the upper limits receive the maximum red intensity. Lower errors are greener and higher errors are redder.

| | MACE Compliant | | | | | | ORB Non-Compliant | | | | | |
|---|---|---|---|---|---|---|---|---|---|---|---|---|
| | Zero-shot | | Fine-tuned | | From scratch | | Zero-shot | | Fine-tuned | | From scratch | |
| Surface | J $m^{-2}$ | % | J $m^{-2}$ | % | J $m^{-2}$ | % | J $m^{-2}$ | % | J $m^{-2}$ | % | J $m^{-2}$ | % |
| $(\bar{1}1\bar{1})$ | 0.47 | 11.86 | 0.11 | 2.71 | 0.03 | 0.68 | 2.07 | 52.86 | 0.29 | 7.47 | 0.07 | 1.69 |
| (110) | 0.66 | 13.04 | 0.08 | 1.53 | 0.11 | 2.09 | 3.38 | 66.31 | 0.34 | 6.62 | 0.10 | 1.90 |
| (100) | 1.08 | 16.20 | 0.23 | 3.39 | 0.05 | 0.69 | 5.78 | 86.56 | 0.43 | 6.38 | 0.12 | 1.85 |
| (111) | 0.47 | 11.86 | 0.11 | 2.71 | 0.03 | 0.68 | 2.21 | 56.47 | 0.29 | 7.46 | 0.10 | 2.54 |
| (101) | 0.87 | 15.09 | 0.16 | 2.80 | 0.19 | 3.32 | 3.46 | 60.01 | 0.37 | 6.39 | 0.03 | 0.57 |
| $(\bar{1}11)$ | 1.31 | 20.06 | 0.36 | 5.54 | 0.51 | 7.72 | 4.68 | 71.42 | 0.25 | 3.83 | 0.05 | 0.77 |
| $(\bar{1}01)$ | 1.41 | 19.89 | 0.23 | 3.20 | 0.33 | 4.61 | 5.79 | 81.76 | 0.43 | 6.08 | 0.10 | 1.47 |
| (001) | 0.90 | 14.85 | 0.10 | 1.70 | 0.23 | 3.83 | 1.65 | 27.22 | 0.32 | 5.20 | 0.08 | 1.23 |
| (011) | 1.38 | 19.64 | 0.28 | 3.99 | 0.45 | 6.47 | 5.23 | 74.61 | 0.48 | 6.85 | 0.13 | 1.88 |
| (211) | 1.22 | 17.48 | 0.28 | 4.01 | 0.41 | 5.91 | 4.92 | 70.68 | 0.36 | 5.20 | 0.07 | 0.97 |
| (010) | 1.45 | 17.58 | 0.19 | 2.30 | 0.05 | 0.61 | 6.01 | 73.14 | 0.33 | 4.04 | 0.04 | 0.45 |

TABLE S8. Slab metadata reconstructed from the first frame of each supplied orientation. Thickness is the atomic extent projected along the surface normal, vacuum is the normal cell height minus that extent, and overlap denotes exact frame matches in the filtered training split.

| Miller index | Atoms | Slab thickness (Å) | Vacuum thickness (Å) | Surface area (Å$^2$) | Exact training overlap |
|---|---|---|---|---|---|
| -11-1 | 60 | 13.818 | 20.000 | 49.450 | Yes (118 frames) |
| 110 | 60 | 16.917 | 20.000 | 38.647 | Yes (158 frames) |
| 100 | 60 | 24.621 | 20.000 | 27.654 | Yes (194 frames) |
| 111 | 60 | 13.818 | 20.000 | 49.450 | No exact frame |
| 101 | 60 | 16.833 | 20.000 | 41.430 | Yes (89 frames) |
| -111 | 60 | 15.611 | 20.000 | 44.460 | Yes (86 frames) |
| -101 | 60 | 19.685 | 20.000 | 35.325 | Yes (92 frames) |
| 001 | 60 | 25.561 | 20.000 | 26.785 | Yes (80 frames) |
| 011 | 60 | 18.410 | 20.000 | 38.030 | Yes (166 frames) |
| 211 | 240 | 19.343 | 20.000 | 141.051 | Yes (10 frames) |
| 010 | 60 | 25.674 | 20.000 | 26.997 | Yes (51 frames) |

was found, although this does not necessarily imply that all of its local atomic environments are outside the training distribution.

## IV. VIBRATIONAL PROPERTIES

All six model/strategy combinations were used on the exact fractional q-points of the Materials Project DFT bands for mp-2858 [5, 6]. Thus, all dispersions in Figure S2 share the Setyawan–Curtarolo Materials Project path. Negative frequencies are retained throughout.

Because direct-force models do not necessarily derive their forces from a scalar potential-energy surface, their applicability to vibrational calculations requires additional examination. Figure S3 compares the zero-shot phonon dispersions obtained with the direct and conservative ORB-v3 variants.

For the DOS comparison, each curve was normalized to unit integrated area. Although the DOS density is non-negative, the associated frequency coordinates were allowed to extend below 0 THz. The normalized curves were then linearly interpolated onto a common 3000-point grid spanning the frequency interval shared by the model and DFT data. The reported integrated absolute error and root-mean-square error are

$$\mathrm{IAE}_{\mathrm{DOS}} = \int |g_{\mathrm{model}}(\omega) - g_{\mathrm{DFT}}(\omega)| \, \mathrm{d}\omega, \tag{S3}$$

$$\mathrm{RMSE}_{\mathrm{DOS}} = \left[ \frac{1}{N_\omega} \sum_{j=1}^{N_\omega} \left( g_{\mathrm{model}}(\omega_j) - g_{\mathrm{DFT}}(\omega_j) \right)^2 \right]^{1/2}. \tag{S4}$$

TABLE S9. Phonon metrics obtained on the common Materials Project q-point path. Band RMSE and MAE compare the sorted frequencies of all 36 modes at all 112 q-points. Maximum-frequency error is the absolute difference between the largest model and DFT frequencies. DOS curves were normalized to unit area and interpolated to a common 3000-point overlap grid. Frequencies and their errors are in THz; DOS RMSE is in $\mathrm{THz}^{-1}$. Cell colors use fixed pastel scales shared by MACE and ORB for each metric: 0–3 THz for band RMSE and MAE, 0–1 for DOS IAE, 0–0.05 $\mathrm{THz}^{-1}$ for DOS RMSE, 0–5 THz for maximum-frequency error, and 0–0.05 for the imaginary-mode fraction. Values equal to or above an upper limit receive the maximum red intensity.

| Model | Strategy | Band RMSE | Band MAE | DOS IAE | DOS RMSE | Max. freq. error | Imaginary fraction |
|---|---|---|---|---|---|---|---|
| MACE | Zero-shot | 0.785 | 0.649 | 0.312 | 0.017 | 0.616 | 0.001 |
| MACE | Fine-tuned | 1.653 | 1.459 | 0.413 | 0.023 | 2.900 | 0.000 |
| MACE | From scratch | 1.598 | 1.450 | 0.447 | 0.025 | 3.624 | 0.000 |
| ORB | Zero-shot | 0.540 | 0.380 | 0.174 | 0.010 | 0.248 | 0.000 |
| ORB | Fine-tuned | 3.067 | 2.825 | 0.557 | 0.034 | 4.587 | 0.000 |
| ORB | From scratch | 10.446 | 9.617 | 0.875 | 0.225 | 19.745 | 0.117 |

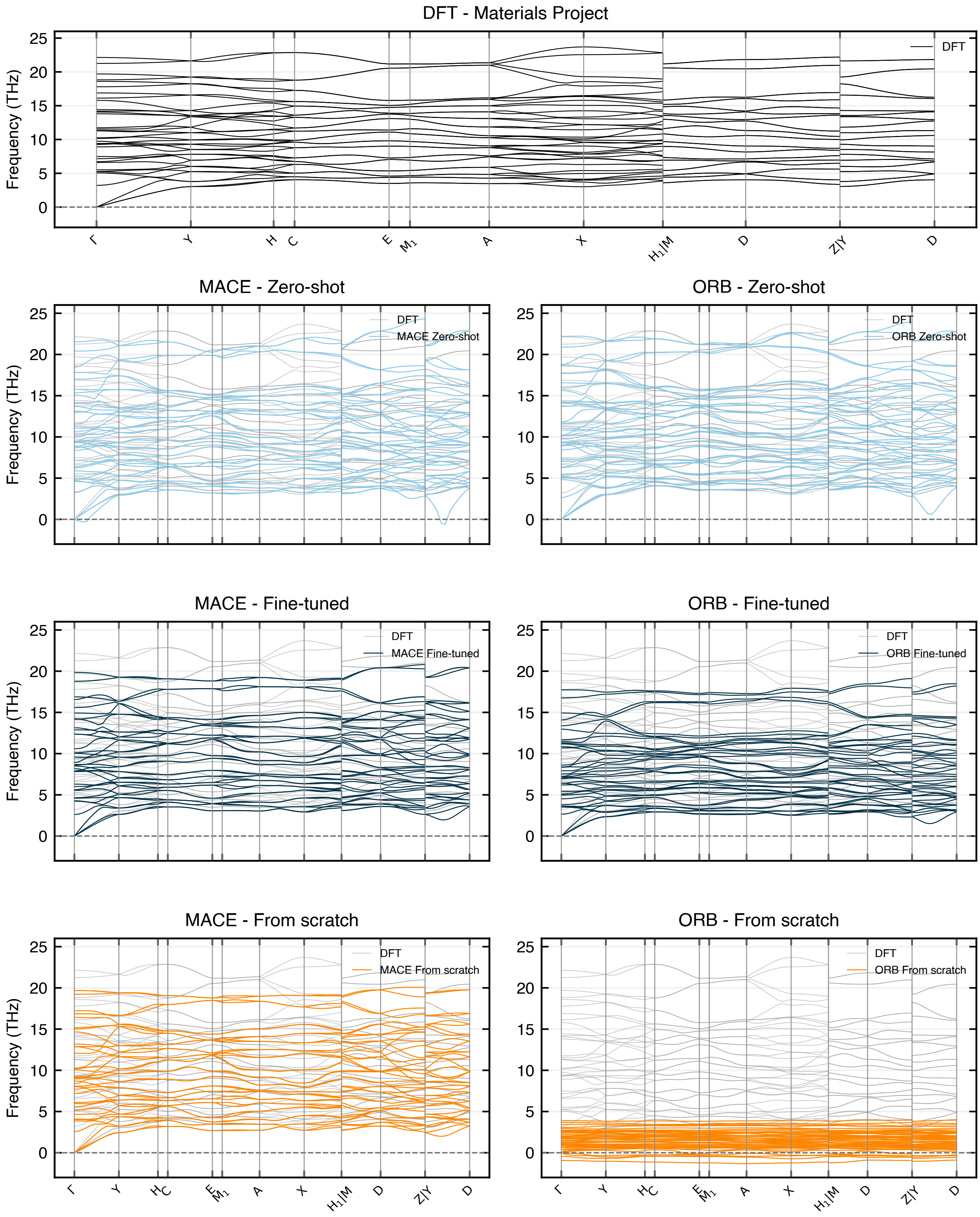


FIG. S2. Materials Project DFT phonon dispersion for mp-2858 (top row, repeated to anchor both columns), followed by MACE (left) and ORB (right) dispersions for zero-shot, fine-tuned, and from-scratch strategies. Every panel uses the same Materials Project path and q-point sampling. Horizontal dashed lines mark zero frequency and vertical lines mark path boundaries.

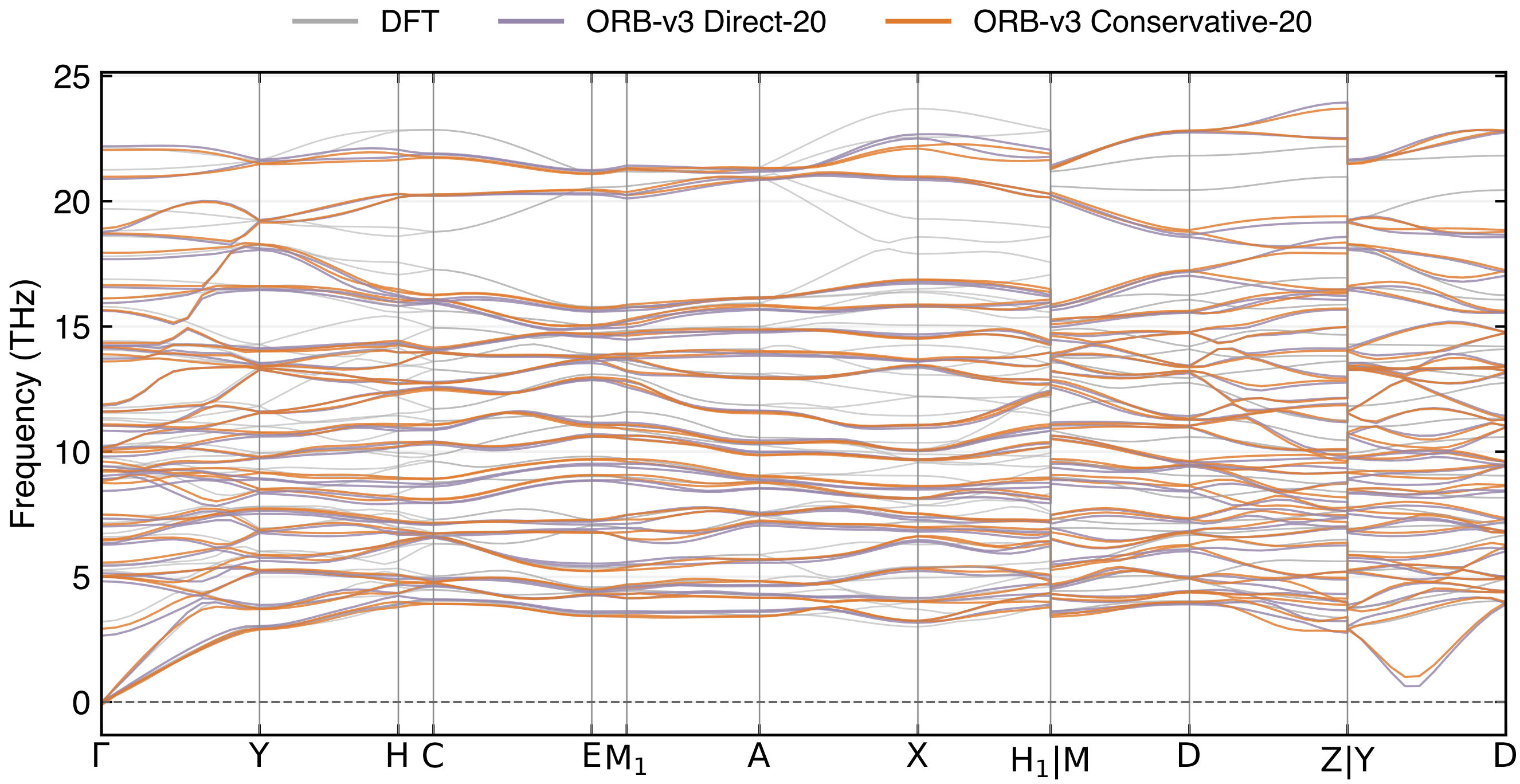


FIG. S3. Comparison of zero-shot phonon dispersions for monoclinic $ZrO_2$ predicted by ORB-v3 Direct-20 and ORB-v3 Conservative-20, with Materials Project DFT as the reference. Horizontal dashed and vertical lines indicate zero frequency and path-segment boundaries, respectively.

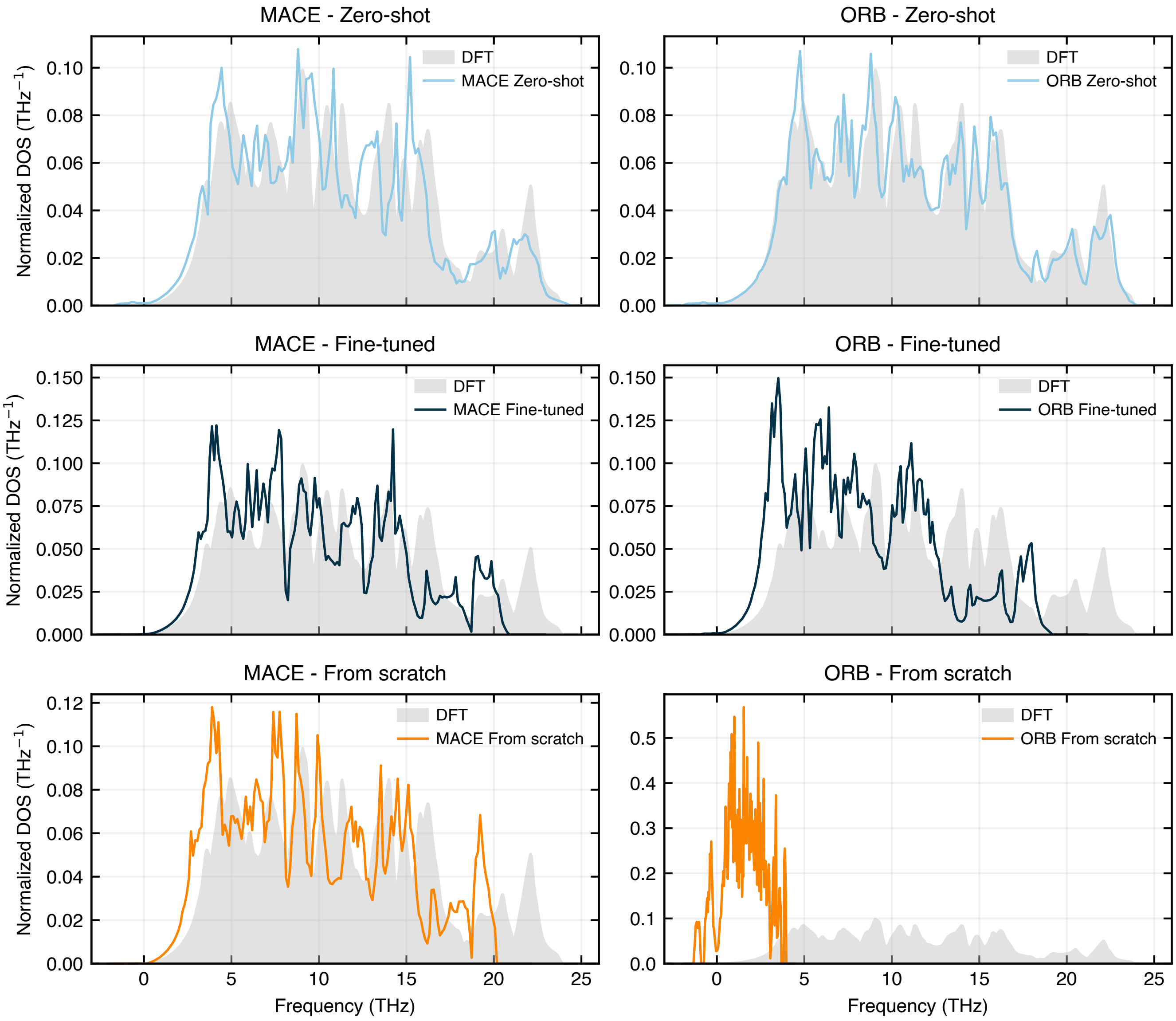


FIG. S4. Unit-area normalized vibrational DOS for MACE (left) and ORB (right), compared with the available DFT DOS. All comparisons use the same interpolation and normalization procedure described in eqs. (S3) and (S4).

## V. ADDITIONAL MOLECULAR-DYNAMICS DIAGNOSTICS

The six neck trajectories were analyzed using the Atomic Simulation Environment (ASE) [7]. Each trajectory spans 42.096 ps and was generated using an integration time step of 1 fs, with three different seeds. The change in total energy relative to the initial configuration was calculated as

$$\Delta E(t) = E(t) - E(0). \tag{S5}$$

The tensile molecular dynamics simulations employed the same $ZrO_2$ neck structure and loading protocol as the DFT AIMD reference [8]. All simulations started directly from the same initial DFT configuration, without additional geometry optimization or atomic constraints.

The MLIP trajectories were propagated under fully periodic boundary conditions in the NVT ensemble at 300 K, using a Nosé–Hoover chain thermostat with a damping time of 100 fs and an integration time step of 1 fs. Following an initial equilibration period of approximately 2 ps, the simulation cell was incrementally elongated along the pulling direction in steps of 0.5 Å. Approximately 2 ps of molecular dynamics was performed between consecutive elongation steps. This procedure resulted in a total elongation of 10 Å after 42.096 ps. At each extension step, the atomic coordinates were scaled consistently with the simulation cell.

The root-mean-square displacements (RMSDs) presented in Figure S5 were calculated relative to the initial configuration after unwrapping atomic trajectories across periodic boundaries and removing overall rigid-body translation and rotation. The minimum Zr–O distance was defined as the shortest Zr–O separation in each frame, accounting for periodic boundary conditions.

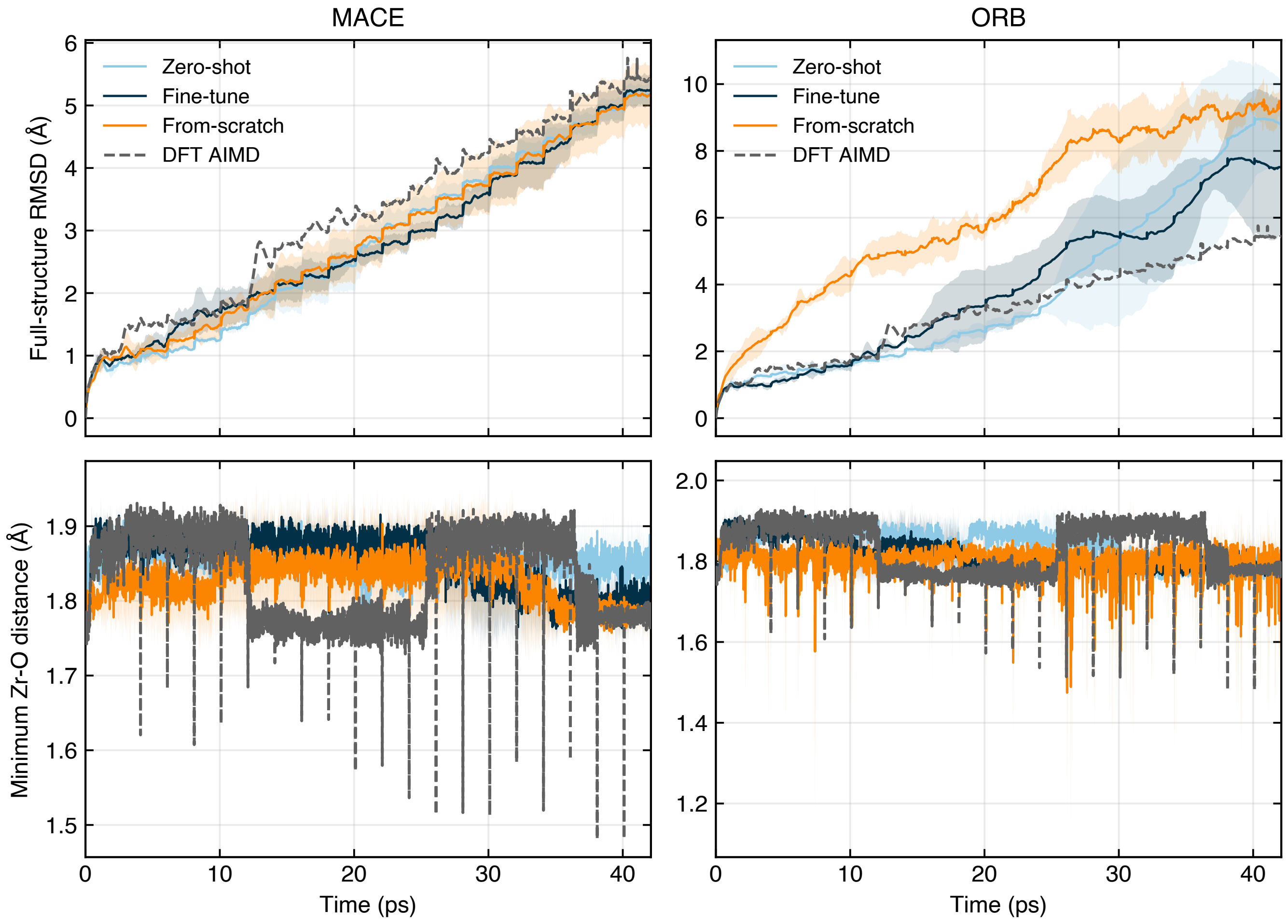


FIG. S5. Full-structure RMSD and minimum periodic Zr–O distance for every trajectory. All RMSDs use the same initial-frame reference and alignment convention.

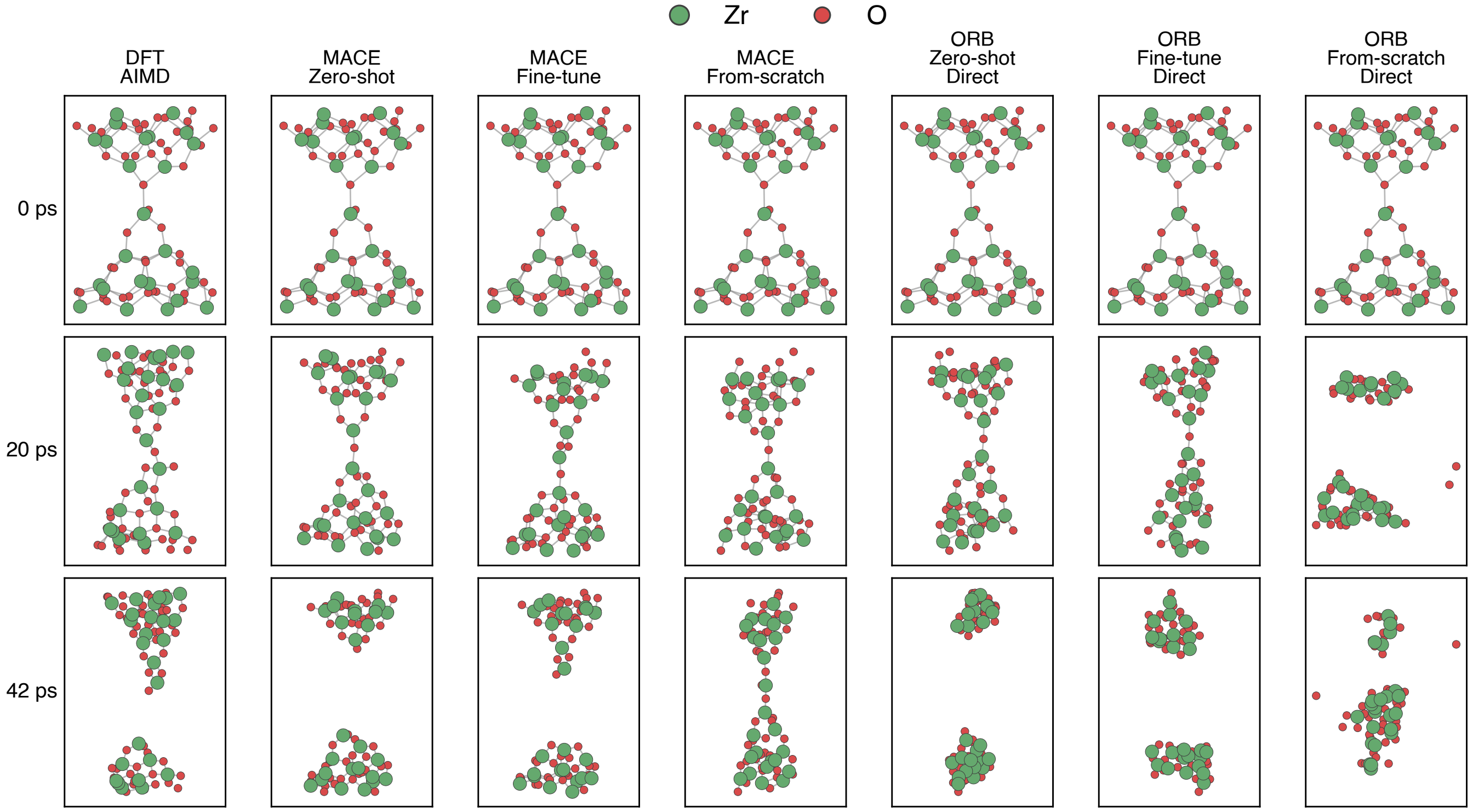


FIG. S6. Representative trajectory snapshots at 0, 10, and 20 ps. Every panel uses the same cropped $x$-$z$ field of view and enlarged element-specific markers, so changes in structural extent remain directly comparable while individual atoms remain visible. Trajectories are neither recentered nor independently wrapped for display.

The energy-trend slope was obtained by fitting the energy per atom as a linear function of simulation time after excluding the initial 2 ps equilibration period. The reported value is the absolute fitted slope in eV atom$^{-1}$ ps$^{-1}$. Under thermostatted tensile loading, this quantity includes thermostat energy exchange and mechanical work.

TABLE S10. Structural diagnostics from the trajectories. RMSDs are measured relative to the first production frame after center removal and mass-unweighted Kabsch alignment. Absolute slopes were fitted to the total energy per atom after excluding the first 2 ps.

| Model | Strategy | Final RMSD (Å) | Maximum RMSD (Å) | Drift slope (eV atom$^{-1}$ ps$^{-1}$) | Minimum Zr–O (Å) |
|---|---|---|---|---|---|
| DFT | AIMD | 5.4292 | 5.7575 | 1.5253e-03 | 1.4810 |
| MACE | Zero-shot | 1.1408 | 1.2993 | 4.0275e-07 | 1.7689 |
| MACE | Fine-tune | 0.4012 | 0.4390 | 5.4444e-08 | 1.7750 |
| MACE | From-scratch | 0.7878 | 0.8730 | 2.5762e-08 | 1.7738 |
| ORB | Zero-shot | 0.5943 | 0.6135 | 1.2248e-03 | 1.6728 |
| ORB | Fine-tune | 1.2025 | 1.2081 | 2.2170e-03 | 1.6745 |
| ORB | From-scratch | 19.2461 | 19.2461 | 3.3010e-03 | 1.5386 |

These trajectories represent a single initial-velocity realization and are used as diagnostic examples rather than a statistical stability benchmark.

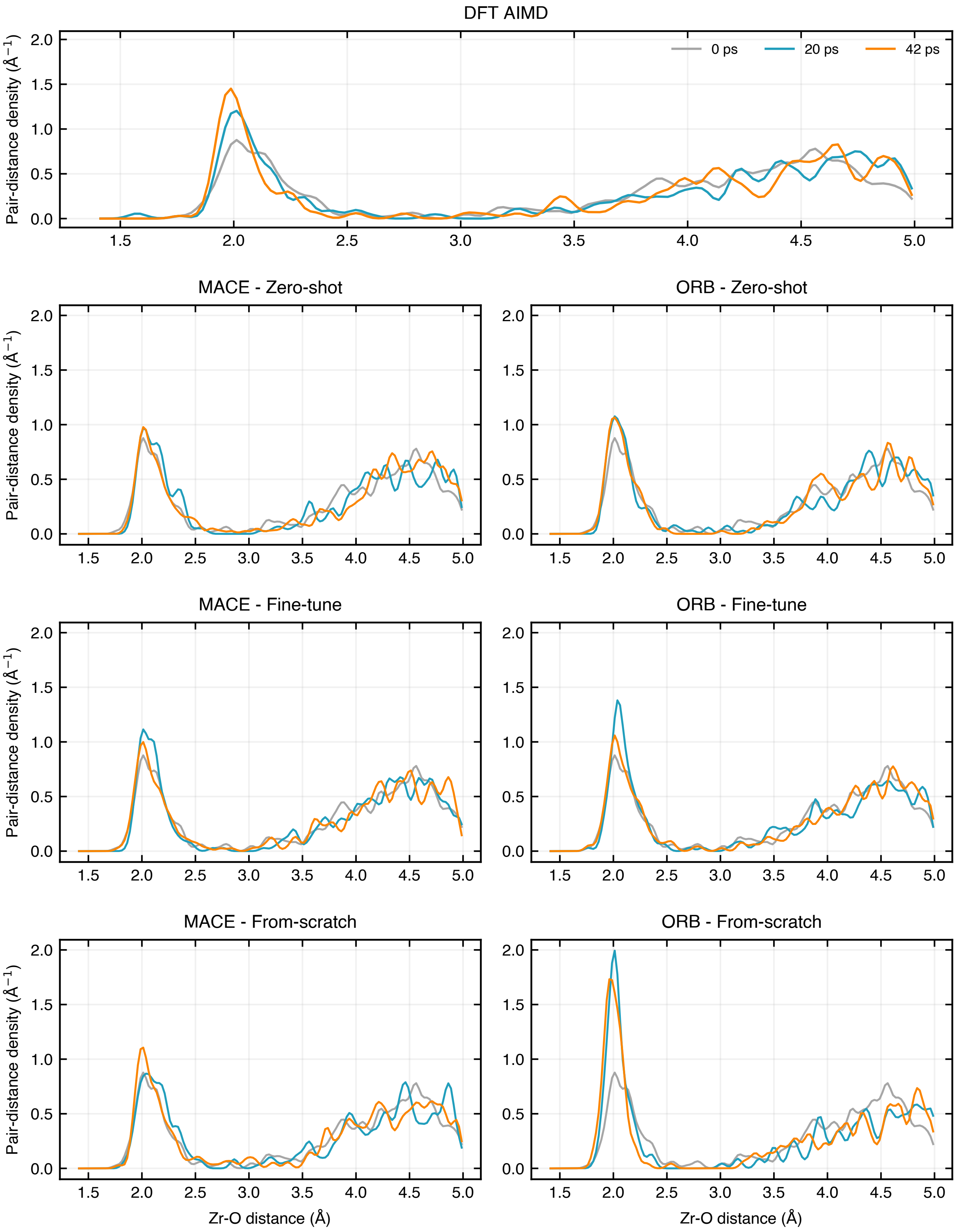


FIG. S7. Zr–O minimum-image pair-distance distributions at 0, 10, and 20 ps. Histograms use a 0.025 Å bin width and light Gaussian smoothing over neighboring bins. This complementary structural fingerprint exposes changes in local and intermediate-range order that may be obscured by projection overlap in the atomic snapshots.



\end{document}